\documentclass[a4paper,prd,notitlepage,aps,11pt,amsmath,amsfonts,amssymb,longbibliography]{revtex4-1}
\usepackage[colorlinks=true]{hyperref}
\usepackage{amsthm}
\usepackage{enumerate}
\usepackage{subfigure}
\usepackage{graphicx}
\usepackage{rotating}
\usepackage{tikz}
\usepackage[utf8x]{inputenc}
\usepackage[T1]{fontenc}

\newtheorem{stmt}{Statement}
\DeclareMathOperator{\sgn}{sgn}

\begin{document}

\title{Dynamical systems approach and generic properties of $f(T)$ cosmology}

\author{Manuel Hohmann}
\email{manuel.hohmann@ut.ee}
\affiliation{Laboratory of Theoretical Physics, Institute of Physics, University of Tartu, W. Ostwaldi 1, 50411 Tartu, Estonia}

\author{Laur J\"arv}
\email{laur.jarv@ut.ee}
\affiliation{Laboratory of Theoretical Physics, Institute of Physics, University of Tartu, W. Ostwaldi 1, 50411 Tartu, Estonia}

\author{Ulbossyn Ualikhanova}
\email{ulbossyn.ualikhanova@ut.ee}
\affiliation{Laboratory of Theoretical Physics, Institute of Physics, University of Tartu, W. Ostwaldi 1, 50411 Tartu, Estonia}

\begin{abstract}
We present a systematic analysis of the dynamics of flat Friedmann-Lema\^{i}tre-Robertson-Walker cosmological models with radiation and dust matter in generalized teleparallel $f(T)$ gravity. We show that the cosmological dynamics of this model are fully described by a function \(W(H)\) of the Hubble parameter, which is constructed from the function \(f(T)\). After reducing the phase space to two dimensions, we derive the conditions on \(W(H)\) for the occurrence of de Sitter fixed points, accelerated expansion, crossing the phantom divide, and finite time singularities. Depending on the model parameters, it is possible to have a bounce (from contraction to expansion) or a turnaround (from expansion to contraction), but cyclic or oscillating scenarios are prohibited. As an illustration of the formalism we consider power law $f(T) = T + \alpha(-T)^n$ models, and show that these allow only one period of acceleration and no phantom divide crossing.
\end{abstract}

\maketitle
%\tableofcontents

\section{Introduction}\label{sec:intro}
General relativity (GR), which relates the effects of gravity to spacetime curvature, has been highly successful in describing a wide range of phenomena. Teleparallel gravity~\cite{Moller:1961,Aldrovandi:2013wha,Maluf:2013gaa} employs torsion instead of curvature, conceptually distinguishes between gravitation and inertia, and builds up its theoretical formulation more in line with gauge theories~\cite{Cho:1975dh,Hayashi:1977jd,Aldrovandi:2013wha}. Despite the difference in mathematical setup and interpretation, teleparallel gravity is equivalent to general relativity in all physical predictions. This follows from the correspondence of the respective field equations, since the curvature scalar $R$ in the Einstein-Hilbert action of general relativity differs from the torsion scalar $T$ in the action of teleparallel equivalent of general relativity (TEGR) only by a total divergence term, $R = -T-2\nabla^\mu T^\lambda_{\phantom{\lambda}\mu\lambda}$, where $T^\lambda_{\phantom{\lambda}\mu\nu}$ are the components of the torsion tensor.

In search for good models to describe the phenomena of dark energy, dark matter, and inflation, many researchers have looked beyond GR, generalizing its Lagrangian to an arbitrary function of curvature, $f(R)$, leading to fourth order field equations~\cite{Capozziello:2010zz,DeFelice:2010aj}. In the same vein the Lagrangian of teleparallel gravity has been generalized to $f(T)$~\cite{Bengochea:2008gz,Linder:2010py}. The ensuing field equations are of second order, and we get a new class of theories essentially different from their counterparts based on curvature.

The original approach to $f(T)$ gravity had a problem that the action failed to be invariant under the local Lorentz transformation of the tetrad fields~\cite{Li:2010cg,Sotiriou:2010mv}. This made the theory subject to preferred frame of reference effects, spurious degrees of freedom, and acausality~\cite{Li:2011rn,Ong:2013qja,Izumi:2013dca,Chen:2014qtl}. The issue can be remedied by realizing that the Weitzenb\"ock connection originally used in teleparallel gravity is not the most general connection consistent with nonzero torsion and vanishing curvature; one can also allow purely inertial spin connection~\cite{Krssak:2015rqa,Krssak:2015lba}. This leads to a covariant approach to $f(T)$ gravity whereby one tackles the field equations by invoking a reference tetrad which encodes the inertial effects~\cite{Krssak:2015oua, Krssak:2017nlv}.

It is remarkable, that while many solutions in $f(T)$ gravity need to be reconsidered in view of the covariant approach, the diagonal tetrad corresponding to flat Friedmann-Lema\^{i}tre-Robertson-Walker (FLRW) universe is already ``proper'' and consistent with the covariant view~\cite{Krssak:2015oua}. Already the first studies of FLRW cosmology in $f(T)$ gravity pointed out the possibility that it can naturally lead to accelerated expansion of the universe without any extra matter component, thus being interesting to model dark energy and inflation~\cite{Bengochea:2008gz,Linder:2010py,Myrzakulov:2010vz,Wu:2010xk}. Later a number of works have focused upon various cosmological aspects of $f(T)$ models, from background evolution and growth of perturbations to comparison with observational data, see Refs.~\cite{Cai:2015emx,Nojiri:2017ncd} for reviews.

The method of dynamical systems is a widely used set of tools in cosmology to obtain a qualitative assessment of the behavior of solutions in a model, without delving into the often almost impossible task to find the analytic form of the solutions. While dynamical systems have been helpful in uncovering the main features of solutions in particular models~\cite{Wu:2010xk,Zhang:2011qp,Jamil:2012yz,Jamil:2012nma,Feng:2014fsa,Cai:2015emx}, there have been only a few papers attempting a more systematic analysis of generic $f(T)$ cosmology~\cite{Zhang:2011qp,Setare:2012ry,Setare:2013xh,Mirza:2017vrk}. Our present study aims at completing this task by deriving the general expressions for de Sitter fixed points, acceleration, phantom dark energy, and finite time singularities. The method and formulas we present can be easily applied to study specific models, or for a heuristic construction of phenomenologically desirable scenarios.

The key in the dynamical systems analysis is choosing suitable dynamical variables, as also highlighted by recent insights in the study of $f(R)$ gravity~\cite{Carloni:2015jla,Alho:2016gzi}. One typically adopts Hubble-rescaled dimensionless variables evolving in dimensionless expansion-logarithmic time parameter $N=\ln a$. This leads to an at first three-dimensional phase space for flat FLRW $f(T)$ cosmology with radiation and dust matter components~\cite{Wu:2010xk,Zhang:2011qp,Jamil:2012nma,Jamil:2012yz,Setare:2012ry,Setare:2013xh,Feng:2014fsa}. There is a further redundancy. In $f(T)$ gravity the Einstein equations are of second order, so in cosmology there are in principle five dynamical quantities $(a, H, T,\rho_r,\rho_m)$. Not all of them are independent, as in the equations the flat FLRW the scale factor $a$ occurs only within the Hubble parameter $H=\frac{\dot{a}}{a}$, the density of radiation $\rho_r$ and dust matter $\rho_m$ are related to $H$ and $f(T)$ by a Friedmann constraint equation, and in addition there is a geometric identity $T=-6H^2$. Therefore the physical phase space is two-dimensional, spanned by two variables given by combinations of the quantities mentioned before. The situation can be contrasted with generic $f(R)$ gravity where the Einstein equations are of fourth order and in cosmology there are six dynamical quantities $(a, H, R, \dot{R},\rho_r,\rho_m)$. Again $a$ is subsumed into $H$, and one quantity can be expressed via others by the Friedmann constraint, but the geometric identity $R=6\dot{H}+12H^2$ does not reduce $R$ to $H$. So the phase space of flat FLRW $f(R)$ cosmology with radiation and dust is four-dimensional~\cite{DeFelice:2010aj,Carloni:2015jla,Alho:2016gzi}.

As in Refs.~\cite{Zhang:2011qp,Feng:2014fsa,Mirza:2017vrk} we reduce the flat FLRW $f(T)$ cosmology phase space to two dimensions. Our choice of the dynamical variables allows a straightforward physical interpretation of results. Taking the Hubble parameter $H$ to be one of the variables makes the fixed points correspond to de Sitter (or Minkowski) spacetime. The second variable given by the ratio of radiation energy density to overall matter energy density makes the flow from radiation to dust matter domination in expanding universe graphic on the other axis. We follow the evolution of the system in basic cosmological time, in order to study both expanding and contracting phases under the same footing.

To illustrate our general results we consider a simple class of models $f(T)=T+\alpha(-T)^n$ as an example. These models allow the cosmic evolution from radiation domination through matter domination to dark energy domination eras~\cite{Bengochea:2008gz,Linder:2010py}, superbounce~\cite{Odintsov:2015uca}, initial singularity crossing~\cite{Awad:2017sau}, future sudden singularities~\cite{Bamba:2012vg}, but no phantom crossing~\cite{Wu:2010mn}. Its fixed points have been studied in Hubble-rescaled variables~\cite{Wu:2010xk,Zhang:2011qp,Setare:2012ry,Feng:2014fsa,Mirza:2017vrk}, and some analytic solutions are also known~\cite{Paliathanasis:2016vsw}. Constraints from various sets of observations can be found in Refs.~\cite{Bengochea:2008gz,Wu:2010mn,Bengochea:2010sg,Wei:2011mq,Nesseris:2013jea,Camera:2013bwa,Geng:2015hen,Basilakos:2016xob,Nunes:2016qyp,Oikonomou:2016jjh,Nunes:2016plz,Malekjani:2016mtm,Farrugia:2016qqe,Capozziello:2017bxm}. We show how these features are reflected with our method and a comprehensive picture emerges.

The outline of the paper is as follows. In section~\ref{sec:action} we briefly review the action and cosmological field equations of \(f(T)\) gravity, and show that they are fully defined in terms of the Friedmann function \(W(H)\). We then cast these equations into the form of a dynamical system in section~\ref{sec:dynsys}, and read off a number of properties of this system: its boundaries and fixed points, as well as the possibility of bounces, turnarounds and oscillating universe solutions. We then discuss finite time singularities in section~\ref{sec:singular}. Observable properties, in particular the accelerating expansion of the universe and the properties of dark energy, are delineated in section~\ref{sec:observ}. In section~\ref{sec:example} we apply our general formalism to a generic power law model and show how its parameters influence the properties of the dynamical system. We end with a conclusion in section~\ref{ses:conclusion}. In order to collect and summarize the results obtained on the physical phase space, we provide a graphical index of phase space points in the appendix~\ref{app:index}.

\section{Action and cosmological field equations}\label{sec:action}
In this section we briefly review the cosmological dynamics of \(f(T)\) gravity, starting from the most general action and cosmological field equations. We show that if we express them through the Hubble parameter of a FLRW spacetime, they take the form of one constraint equation and one dynamical equation. We further display these field equations for the special case that the matter content of the universe is constituted by both dust matter and radiation, and show their consistency with the corresponding continuity equations for this choice of the matter content.

The starting point of our derivation is the action functional of \(f(T)\) gravity~\cite{Bengochea:2008gz,Linder:2010py,Aldrovandi:2013wha,Cai:2015emx},
\begin{equation}\label{eqn:action}
S = \frac{1}{16\pi G}\int\left|e\right|f(T)d^4x\,,
\end{equation}
with an arbitrary function \(f(T)\) of the torsion scalar
\begin{equation}\label{eqn:torsionscal}
T = \frac{1}{4}T^{\rho}{}_{\mu\nu}T_{\rho}{}^{\mu\nu} + \frac{1}{2}T^{\rho}{}_{\mu\nu}T^{\nu\mu}{}_{\rho} - T^{\mu}{}_{\rho\mu}T^{\nu\rho}{}_{\nu}\,.
\end{equation}
The dynamical variable is given by the tetrad field \(e^i_{\mu}\), in terms of which the torsion tensor is expressed as
\begin{equation}
T^{\rho}{}_{\mu\nu} = \Gamma^{\rho}{}_{\nu\mu} - \Gamma^{\rho}{}_{\mu\nu} = e_i^{\rho}\left(\partial_{\mu}e^i_{\nu} - \partial_{\nu}e^i_{\mu} + \omega^i{}_{j\mu}e^j_{\nu} - \omega^i{}_{j\nu}e^j_{\mu}\right)\,,
\end{equation}
where the flat spin connection \(\omega^i{}_{j\mu}\) is introduced in order to render the theory covariant under local Lorentz transformations. For our cosmological setting, we assume a flat Friedmann-Lema\^{i}tre-Robertson-Walker (FLRW) universe, for which we can choose the tetrad to be
\begin{equation}
e^i_{\mu} = \mathrm{diag}(1, a, a, a)
\end{equation}
with the usual cosmological scale factor \(a = a(t)\). One finds that in this case the tetrad is ``proper'' and the inertial spin connection vanishes~\cite{Krssak:2015oua}. The torsion scalar reduces to
\begin{equation}\label{eqn:geomcons}
T = -6\frac{\dot{a}^2}{a^2} = -6H^2\,,
\end{equation}
where \(H\) is the Hubble parameter. From our assumption of cosmological symmetry, i.e., homogeneity and isotropy, further follows that the matter energy-momentum tensor must take the form of a perfect fluid,
\begin{equation}
T^{\mu\nu} = (\rho + p)u^{\mu}u^{\nu} + pg^{\mu\nu}\,,
\end{equation}
where $\rho$ stands for energy density, $p$ denotes pressure, and the four-velocity
\begin{equation}
u^{\mu} = \partial_t
\end{equation}
is normalized by the metric \(g_{\mu\nu} = \eta_{ij}e^i_{\mu}e^j_{\nu}\). From the action~\eqref{eqn:action} then follow the cosmological field equations~\cite{Bengochea:2008gz,Linder:2010py}
\begin{subequations}
\begin{align}
12H^2f_T + f &= 16\pi G\rho\,,\\
48H^2\dot{H}f_{TT} - (12H^2 + 4\dot{H})f_T - f &= 16\pi Gp\,,
\end{align}
\end{subequations}
where subscripts denote derivatives, i.e.,
\begin{equation}
f_T = \frac{df}{dT}\,, \quad
f_{TT} = \frac{d^2f}{dT^2}\,.
\end{equation}
In the remainder of this article, without a loss of generality we will write \(f(T) = T + F(T)\). In this parametrization the cosmological field equations read
\begin{subequations}
\begin{align}
6H^2 + 12H^2F_T + F &= 16\pi G\rho\,,\\
4\dot{H}(12H^2F_{TT} - F_T - 1) &= 16\pi G(\rho + p)\,.
\end{align}
\end{subequations}
Note that if $F=0$ these equations reduce to the usual Friedmann equations in TEGR and GR. Also note, that if $F_T=\frac{F}{2T}$, i.e, $F \sim \sqrt{-T}$, the cosmological equations are still identical to the TEGR and GR case. However, the full field equations receive corrections, which possibly influence the evolution of perturbations of the cosmological background.

In order to discuss solutions to these field equations, we finally also need an equation of state for the matter content. Here we will use the simple assumption that the matter content is constituted by two components: dust and radiation. The density and pressure thus take the form
\begin{equation}
\rho = \rho_m + \rho_r\,, \quad
p = p_m + p_r\,,
\end{equation}
where the equation of state is given by
\begin{equation}
p_m = 0\,, \quad
p_r = \frac{1}{3}\rho_r\,.
\end{equation}
From these relations follow the matter continuity equations
\begin{equation}\label{eqn:contin}
\dot{\rho}_m = -3H\rho_m\,, \quad
\dot{\rho}_r = -4H\rho_r\,.
\end{equation}
The cosmological field equations then finally take the form
\begin{subequations}\label{eqn:friedmann}
\begin{align}
W &= 16\pi G(\rho_m + \rho_r)\,,\label{eqn:friedcons}\\
-\dot{H}\frac{W_H}{3H} &= 16\pi G\left(\rho_m + \frac{4}{3}\rho_r\right)\,.\label{eqn:frieddyn}
\end{align}
\end{subequations}
Here we have introduced the Friedmann function
\begin{equation}
W(H) = F + 6H^2 + 12H^2F_T\,,
\end{equation}
keeping in mind the relation~\eqref{eqn:geomcons} between $T$ and $H$, and where the subscript stands for differentiation $W_H = dW/dH$. As we will see in the following, the function $W(H)$ encodes the main cosmological features of any given $f(T)$ gravity model. The equations~\eqref{eqn:contin} and~\eqref{eqn:friedmann} are the basis of the current study. Note that they are not independent of each other. For our analysis, we have to remove this redundancy and apply the constraint equation, in order to obtain an unconstrained dynamical system. This will be done in the following section.

\section{Dynamical systems approach}\label{sec:dynsys}
We will now cast the cosmological equations of motion~\eqref{eqn:friedmann} and~\eqref{eqn:contin} detailed in the previous section into the language of dynamical systems and derive some of its properties. We start by determining the phase space of the dynamical system and its evolution equations in section~\ref{ssec:phase}. Since this phase space will turn out to be unbounded in one direction, we perform a coordinate transformation in section~\ref{ssec:dynvary}, which maps the system into a compact region. We then discuss three particular features of the dynamical system: its fixed points in section~\ref{ssec:fix}, the possibility of crossing the line \(H = 0\) in section~\ref{ssec:bounce} and the possibility of an oscillating universe in section~\ref{ssec:cyclic}.

\subsection{Phase space and evolution equations}\label{ssec:phase}
In the previous section we have seen that the Hubble parameter \(H\) and the energy densities \(\rho_m\) and \(\rho_r\) are not independent due to the algebraic Friedmann constraint~\eqref{eqn:friedcons}. The physical phase space of our dynamical system is thus a hypersurface of codimension 1 in the space
\begin{equation}\label{eqn:physphase}
\{(H, \rho_m, \rho_r) \quad | \quad
H \in (-\infty, \infty)\,, \quad
\rho_m \in [0,\infty)\,, \quad
\rho_r \in [0,\infty)\}\,,
\end{equation}
which we parametrize as follows. We introduce a new variable
\begin{equation}\label{eqn:xdef}
X=\frac{\rho_r}{\rho_r+\rho_m}
\end{equation}
for the ratio of radiation to the total energy density. The original densities are then recovered as
\begin{equation}
\rho_r=X(\rho_r+\rho_m) \,,\quad
\rho_m=(1-X)(\rho_r+\rho_m) \,,
\end{equation}
where the total energy density on the right hand side is determined in terms of \(H\) by the Friedmann constraint~\eqref{eqn:friedcons}. One easily reads off that the physical phase space is restricted to $H \in (-\infty, \infty), X \in [0,1]$. Another bound follows from the Friedmann constraint and the validity of the null energy condition, which implies that the total matter energy density \(\rho_m + \rho_r\) must be non-negative and finite. From the Friedmann constraint~\eqref{eqn:friedcons} follows that this is equivalent to \(W(H) \geq 0\). The physical phase space is thus finally given by
\begin{equation}\label{eqn:phaseh}
\mathcal{P} = \{(H,X) \quad | \quad -\infty < H < \infty\,, \quad 0 \leq X \leq 1\,, \quad 0 \leq W(H) < \infty\}\,.
\end{equation}
We now discuss the dynamics of the new variables. Taking the time derivative of the definition~\eqref{eqn:xdef} and using the continuity equations~\eqref{eqn:contin} one finds
\begin{equation}\label{eqn:hdynx}
\dot{X} = HX(X - 1)\,.
\end{equation}
Similarly, we can solve the Friedmann equation~\eqref{eqn:friedcons} for \(\dot{H}\) and use the definition~\eqref{eqn:xdef} to obtain
\begin{equation}\label{eqn:hdynh}
\dot{H} = -(X + 3)H\frac{W}{W_H} = -\frac{(X + 3)H}{(\ln W)_H}\,,
\end{equation}
where we have tacitly assumed that the factor \(W_H/H\) by which we divided is nonzero and finite, since otherwise the division would be ill-defined, or defined only via a suitable limiting procedure. We have to keep this in mind later when we will be discussing these limiting cases. The equations~\eqref{eqn:hdynx} and~\eqref{eqn:hdynh} define our dynamical system. Note that our choice for the dynamical variables to describe the two-dimensional phase space is different from Refs.~\cite{Zhang:2011qp,Feng:2014fsa,Mirza:2017vrk}.

We also remark that our model includes the two special cases in which the matter content is given by pure dust matter, \(\rho_r = 0\), and pure radiation, \(\rho_m = 0\). These are obtained by restricting the phase space to \(X = 0\) or \(X = 1\), respectively. From the continuity equation in the form~\eqref{eqn:hdynx} follows that any dynamics that start on these subspaces will also remain there, so that they can be treated as dynamical systems on their own. This will be considered next, when we compactify the phase space and discuss its boundary.

\subsection{Compactified phase space and its boundary}\label{ssec:dynvary}
For various applications, such as drawing phase diagrams and discussing limiting points of trajectories, it is more convenient if the dynamical system is defined on a compact domain. In order to map our dynamical system into a compact domain, we replace the Hubble parameter \(H\) with a new variable \(Y\), which we define such that
\begin{equation}
Y = \frac{H}{\sqrt{1 + H^2}} \quad \Leftrightarrow \quad H = \frac{Y}{\sqrt{1 - Y^2}}\,.
\end{equation}
The domain of these variables in given by \(Y \in (-1, 1), X \in [0, 1]\). We then express the dynamical equations in terms of the new variables, which yields
\begin{subequations}
\begin{align}
\dot{Y} &= -\frac{(X + 3)Y(1 - Y^2)W}{W_H}\,,\\
\dot{X} &= \frac{XY(X - 1)}{\sqrt{1 - Y^2}}\,.
\end{align}
\end{subequations}
This rescaled system will be used later when we show phase diagrams for particular functions \(W\).

Using the new variables \((Y, X)\) we can now discuss the boundary \(\partial\mathcal{P}\) of the phase space. This is important since there may exist trajectories, whose limiting points are located on the boundary, and which may or may not be part of the physical phase space~\eqref{eqn:physphase}. This concerns in particular points at which \(H\) or \(W\) (and hence the total matter energy density) diverges, so that the Friedmann equations~\eqref{eqn:friedmann} become singular, and thus invalid. From the definition~\eqref{eqn:phaseh} of the physical phase space follows that we have the following components of the boundary \(\partial\mathcal{P}\):

\begin{enumerate}[i.]
\item\label{it:bndxz}
The boundary along the line \(X = 0\) belongs to the physical phase space. It contains those points of the phase space where the radiation energy density \(\rho_r\) vanishes, and the matter content of the universe is given by dust matter, or we have a vacuum. From the dynamical equation~\eqref{eqn:hdynx} follows that \(\dot{X}\) vanishes for \(X = 0\), so that any trajectories starting from this boundary stay on the boundary. Later in section~\ref{ssec:fix} we see that regular de Sitter fixed points can reside on this boundary.

\item\label{it:bndxo}
The same properties hold for the boundary along the line \(X = 1\), which contains those points of the phase space where the dust energy density \(\rho_m\) vanishes, and the matter content of the universe is given by radiation, or we have a vacuum. Later in section~\ref{ssec:fix} we see that regular de Sitter fixed points can reside also on this boundary.

\item\label{it:bndwz}
Lines with constant \(H = H^{\star}\) and \(X \in [0, 1]\), where \(W(H^{\star}) = 0\) and the sign of \(W\) is different for $H < H^{\star}$ and $H > H^{\star}$, also correspond to the boundaries, which are part of the physical phase space. They represent vacuum solutions of the field equations. Note that in this case the whole line \(H = H^{\star}, X \in [0, 1]\) represents a single point \((H, \rho_m, \rho_r) = (H^{\star}, 0, 0)\) of the original phase space. These boundaries always contain regular de Sitter fixed points, as we will see in a deeper discussion in section~\ref{ssec:fix}.

\item\label{it:bndwi}
Similarly, also lines with constant \(H = H^{\star}\) and \(X \in [0, 1]\), where \(W \to \infty\), are boundaries. In contrast to the previous case, they do not belong to the physical phase space, since the total matter energy density \(\rho_m + \rho_r\) diverges. However, as we will see in section~\ref{ssec:fix}, there are trajectories which approach these boundaries in infinite time, and they contain points which can be regarded as (singular) fixed points.

\item\label{it:bndhi}
Finally, also the lines \(Y = \pm 1, X \in [0, 1]\) corresponding to \(H \to \pm\infty\) can be treated as boundaries of the (compactified) phase space, provided that \(W\) is non-negative in the corresponding limit. These boundaries are relevant as they may contain limit points of trajectories that are reached in finite time, which correspond to particular types of singularities as discussed in section~\ref{ssec:singii}.
\end{enumerate}

Note that it is also possible that \(\dot{H}\) diverges, for example, at points \((H, X)\) where \(W_H = 0\), and that in this case trajectories reaching this point cannot be continued. However, these can still be regarded as parts of the physical phase space, if the physical quantities \((H, \rho_m, \rho_r)\) constituting the original phase space remain finite. These points are known as sudden singularities and discussed in detail in section~\ref{ssec:singfi}.

Since the boundary \(\partial\mathcal{P}\) may contain a number of interesting points as briefly mentioned above and further discussed in the remaining sections of this article, it will turn out to be more convenient to study the dynamics on the compactified phase space \(\bar{\mathcal{P}} = \mathcal{P} \cup \partial\mathcal{P}\).

\subsection{Fixed points and their stability}\label{ssec:fix}
We now come to the discussion of fixed points of the dynamical system defined by equations~\eqref{eqn:hdynx} and~\eqref{eqn:hdynh} in section~\ref{ssec:phase}. Recall that the fixed points of a dynamical system are points $(H^{\star}, X^{\star})$ in its phase space at which the flow of the dynamics vanishes. For the dynamical system we consider here this amounts to the conditions $\dot{X}=0$ and $\dot{H}=0$. From equation~\eqref{eqn:hdynx} one easily reads off that \(\dot{X} = 0\) if either \(X = 0\), \(X = 1\) or \(H = 0\). The condition for \(\dot{H} = 0\) given by equation~\eqref{eqn:hdynh} requires a more careful treatment, as it depends on the Friedmann function \(W(H)\). We can distinguish the following cases:
\begin{itemize}
\item
For \(W \to 0\) it is obvious that \(\dot{H} \to 0\) in the case that \(W_H\) remains finite. However, also in the case that \(W_H\) either vanishes or diverges for \(W = 0\) we obtain a fixed point. To see this, note that \(\ln W \to -\infty\) when \(W \to 0\). However, this implies that also \((\ln W)_H \to \pm\infty\), where the sign depends on the direction of the limit. Hence, from \(W \to 0\) always follows \(\dot{H} \to 0\). From the Friedmann constraint~\eqref{eqn:friedcons} further follows that the energy density of dust matter and radiation vanishes, so that these fixed points correspond to vacuum solutions. As we will discuss in section~\ref{ssec:efflambda}, they are de Sitter vacuum solutions for \(H^{\star} \neq 0\) and Minkowski vacuum solutions for \(H^{\star} = 0\), which follows from the fact that for a constant Hubble parameter \(\dot{a}/a = H = H^{\star}\) the scale factor behaves as
\begin{equation}\label{eqn;expscalef}
a(t) \sim \exp(H^{\star}t)\,,
\end{equation}
and so is constant or exponentially increasing / decreasing, depending on the sign of \(H^{\star}\).

\item
Following the same line of argumentation, we can also consider the case that \(W \to \infty\) for a finite value of \(H^{\star}\). In this case we find \(\ln W \to \infty\), which again implies \((\ln W)_H \to \pm\infty\), so that also in this case \(\dot{H} \to 0\) and we obtain a fixed point. Note that these points are not part of the physical phase space \(\mathcal{P}\) defined by equation~\eqref{eqn:phaseh}, but lie on the boundary \(\partial\mathcal{P}\). We call them singular fixed points, in order to distinguish them from regular fixed points which belong to the physical phase space. Here for $H^{\star} < 0$ the radiation and dust matter content get compressed to infinite density after infinite time in the future. Analogously for $H^{\star} > 0$ the radiation and matter started from an infinite density state infinite time ago in the past. Note that also in this case the scale factor approaches asymptotically the exponential behavior~\eqref{eqn;expscalef} in the vicinity of the fixed point.

\item
We are left with the case that \(W \to W^{\star} > 0\) remains finite. In this case we can still obtain a fixed point of the equation~\eqref{eqn:hdynh} if \(W_H \to \pm\infty\) diverges. This condition is necessary and sufficient for \(H \neq 0\). For \(H = 0\) the weaker condition that \(W_H/H\) diverges is both necessary and sufficient. However, if we map this fixed point into the original phase space with variables \((H, \rho_m, \rho_r)\), we find the paradox situation that \(\rho_m + \rho_r > 0\) is constant, since \(W = W^{\star} > 0\) is also constant at a fixed point, though in general \(H = H^{\star} \neq 0\). This contradicts the continuity equations~\eqref{eqn:contin}. The reason for this contradiction is the fact that we divided the dynamical equation~\eqref{eqn:frieddyn} by an infinite quantity, and hence generated a previously non-existing solution. However, these points are still relevant in a suitable limit, since they turn out to correspond to a certain class of finite time singularities, as shown in section~\ref{ssec:singff}.
\end{itemize}
Note that the existence and number of fixed points satisfying these conditions depends on the Friedmann function \(W\), and hence on the choice of the function \(F\). We will not make such a choice here and discuss the generic properties of the aforementioned fixed points.

In order to discuss the stability of a fixed point \((H^{\star}, X^{\star})\), we introduce small perturbations \((h, x)\) around the fixed point such that
\begin{equation}
H = H^{\star} + h\,, \quad X = X^{\star} + x\,,
\end{equation}
and then linearise the dynamical equations in \(h\) and \(x\). This leads to a linear system of the form
\begin{equation}
\left(\begin{array}{c}
\dot{h}\\
\dot{x}
\end{array}\right) = J \cdot \left(\begin{array}{c}
h\\
x
\end{array}\right)\,, \quad J = \left.\begin{pmatrix}
\frac{\partial\dot{H}}{\partial H} & \frac{\partial\dot{H}}{\partial X}\\
\frac{\partial\dot{X}}{\partial H} & \frac{\partial\dot{X}}{\partial X}
\end{pmatrix}\right|_{H = H^{\star}, X = X^{\star}}\,,
\end{equation}
where the partial derivatives are given by
\begin{subequations}
\begin{align}
\frac{\partial\dot{H}}{\partial H} &= -(X + 3)\frac{WW_H + HW_H^2 - HWW_{HH}}{W_H^2}\,, &
\frac{\partial\dot{H}}{\partial X} &= -H\frac{W}{W_H}\,,\\
\frac{\partial\dot{X}}{\partial H} &= X(X - 1)\,, &
\frac{\partial\dot{X}}{\partial X} &= (2X - 1)H\,.
\end{align}
\end{subequations}
It follows immediately that \(\partial\dot{H}/\partial X = 0\) whenever \(\dot{H} = 0\). A detailed treatment is necessary for \(\partial\dot{H}/\partial H\). For this purpose we write \(W\) in the form
\begin{equation}
W \approx W^{\star} + c|H - H^{\star}|^b
\end{equation}
in the vicinity of the critical value \(H^{\star}\), where \(W^{\star}, c, b\) are constants. This approximation covers all cases we mentioned before, and we find the following behavior:
\begin{itemize}
\item
To study the case \(W \to 0\), we set \(W^{\star} = 0\) and \(b > 0\). Note that we have \(W_H \to \pm\infty\) for \(b < 1\) and \(W_H \to 0\) for \(b > 1\), while \(W_H\) remains finite for \(b = 1\). In all three cases we find
\begin{equation}\label{eqn:wfixedp}
\dot{H}_H^{\star} = \lim_{H \to H^{\star}}\frac{\partial\dot{H}}{\partial H} = -\frac{(X + 3)H^{\star}}{b}\,.
\end{equation}

\item
To model the case \(W \to \infty\), we consider \(b < 0\), and can likewise set \(W^{\star} = 0\). We find the same limit~\eqref{eqn:wfixedp}.

\item
Finally, we consider the case \(W^{\star} > 0\) with \(b > 0\). If \(H^{\star} \neq 0\), \(W_H\) must diverge in order to obtain a fixed point. This is the case for \(b < 1\). However, in this case \(\partial\dot{H}/\partial H\) also diverges, so that a linear approximation cannot be used to determine the stability of the fixed point, and one must explicitly study the behavior of \(\dot{H}\) near the fixed point. If \(H^{\star} = 0\), we only need \(b < 2\) in order to obtain a fixed point. For \(1 < b < 2\) we find that \(\partial\dot{H}/\partial H\) likewise diverges, while for \(0 < b < 1\) we obtain \(\partial\dot{H}/\partial H \to 0\), and so also in these cases the linearised system is not sufficient. Finally, a special case is given by \(H^{\star} = 0\) and \(b = 1\), in which we find
\begin{equation}\label{eqn:wfixedp2}
\dot{H}_H^{\star} = \lim_{H \to H^{\star}}\frac{\partial\dot{H}}{\partial H} = -\frac{(X + 3)W^{\star}}{c}\,.
\end{equation}
\end{itemize}
From this analysis follows that the Jacobi matrix simplifies significantly at a fixed point, so that
\begin{equation}
J = \begin{pmatrix}
\dot{H}_H^{\star} & 0\\
X^{\star}(X^{\star} - 1) & (2X^{\star} - 1)H^{\star}
\end{pmatrix}\,, \quad \dot{H}_H^{\star} = -A(X^{\star} + 3)\,,
\end{equation}
where the constant \(A\) follows from either the formula~\eqref{eqn:wfixedp} or~\eqref{eqn:wfixedp2}, depending on the nature of the fixed point. Note that \(A > 0\) in the cases \(W \to 0\) at \(H^{\star} > 0\), \(W \to \infty\) at \(H^{\star} < 0\) and \(W > 0\), \(W_H > 0\) at \(H^{\star} = 0\). Similarly, we find \(A < 0\) in the cases \(W \to 0\) at \(H^{\star} < 0\), \(W \to \infty\) at \(H^{\star} > 0\) and \(W > 0\), \(W_H < 0\) at \(H^{\star} = 0\). In these cases a linear approximation is sufficient in order to determine the stability of the fixed points. We thus calculate the eigenvalues of \(J\) for these cases only, and distinguish between the three different conditions obtained from \(\dot{X} = 0\):
\begin{itemize}
\item
\(H^{\star} = 0\): The Jacobi matrix and eigenvalues reduce to
\begin{equation}
J = \begin{pmatrix}
-A(X^{\star} + 3) & 0\\
X^{\star}(X^{\star} - 1) & 0
\end{pmatrix}\,, \quad
\lambda_1 = -A(X^{\star} + 3)\,, \quad
\lambda_2 = 0\,.
\end{equation}
Note that one eigenvalue vanishes, whose corresponding eigenvector is given by \(\partial_X\). This relates to the fact that in this case all points with \(H = 0\) and \(0 \leq X \leq 1\) are non-isolated fixed points. The stability is determined by the remaining eigenvalue, which is positive for \(W_H < 0\), which yields a repeller, and negative for \(W_H > 0\), which yields an attractor.

\item
\(X^{\star} = 0\): The Jacobi matrix and eigenvalues are given by
\begin{equation}
J = \begin{pmatrix}
-3A & 0\\
0 & -H^{\star}
\end{pmatrix}\,, \quad
\lambda_1 = -H^{\star}\,, \quad
\lambda_2 = -3A\,.
\end{equation}
Both eigenvalues are negative for \(W \to 0\) at \(H^{\star} > 0\), which yields an attractor, and positive for \(W \to 0\) at \(H^{\star} < 0\), which yields a repeller. For \(W \to \infty\) the eigenvalues have opposite signs, so that we find a saddle point.

\item
\(X^{\star} = 1\): In this case the Jacobi matrix and eigenvalues take the form
\begin{equation}
J = \begin{pmatrix}
-4A & 0\\
0 & H^{\star}
\end{pmatrix}\,, \quad
\lambda_1 = H^{\star}\,, \quad
\lambda_2 = -4A\,.
\end{equation}
Now the situation is reversed compared to the previous case. Both eigenvalues are negative for \(W \to \infty\) at \(H^{\star} < 0\), which yields an attractor, and positive for \(W \to \infty\) at \(H^{\star} > 0\), which yields a repeller. For \(W \to 0\) the eigenvalues have opposite signs, so that we find a saddle point.
\end{itemize}
In all other cases the linearised analysis is not sufficient, and the sign of \(\dot{H}\) must be studied explicitly in the vicinity of the fixed point using the full, non-linear equations of motion. We can summarize our findings as follows. We start with a classification of regular fixed points, which are elements of the physical phase space \(\mathcal{P}\):
\begin{stmt}\label{stmt:fixreg}
A point \((H^{\star}, X^{\star})\) is a regular fixed point of the dynamical system, if it satisfies one of the following criteria:
\begin{enumerate}[i.]
\item\label{it:fpregisoat}
In the case \(X^{\star} = 0\), \(H^{\star} > 0\), \(W^{\star} = 0\) it is an isolated attractor. The corresponding solution is an expanding de Sitter vacuum solution with scale factor~\eqref{eqn;expscalef}.
\item\label{it:fpregisore}
In the case \(X^{\star} = 0\), \(H^{\star} < 0\), \(W^{\star} = 0\) it is an isolated repeller. The corresponding solution is a contracting de Sitter vacuum solution with scale factor~\eqref{eqn;expscalef}.
\item\label{it:fpregisosa}
In the case \(X^{\star} = 1\), \(H^{\star} \neq 0\), \(W^{\star} = 0\) it is an isolated saddle point. The corresponding solution is physically equivalent to either of the two aforementioned cases.
\item\label{it:fpregnisoat}
Points with \(0 \leq X^{\star} \leq 1\), \(H^{\star} = 0\), \(W^{\star} > 0\) and \(W_H^{\star} > 0\) are non-isolated attractors. The corresponding solution is a static universe with Minkowski geometry, but non-vanishing matter content.
\item\label{it:fpregnisore}
Points with \(0 \leq X^{\star} \leq 1\), \(H^{\star} = 0\), \(W^{\star} > 0\) and \(W_H^{\star} < 0\) are non-isolated repellers. The corresponding solution is a static universe as in the aforementioned case.
\item\label{it:fpregindet}
Fixed points, whose stability cannot be determined from a linearised analysis, are given by:
\begin{enumerate}[a.]
\item\label{it:fpregidwz}
\(H^{\star} = 0\) and \(W^{\star} = 0\); this is a Minkowski vacuum solution.
\item\label{it:fpregidwhd}
\(H^{\star} = 0\), \(W^{\star} > 0\) and \(W_H^{\star}\) diverges; also this is a static universe with non-vanishing matter content.
\item\label{it:fpregidwhz}
\(H^{\star} = 0\), \(W^{\star} > 0\) and \(W_H^{\star} = 0\) such that \(H/W_H \to 0\); this case corresponds to a finite time singularity of type~\ref{it:singtype4}, as shown in section~\ref{ssec:singff}.
\end{enumerate}
\end{enumerate}
\end{stmt}
There are a number of fixed points, which either lie outside the physical phase space~\eqref{eqn:physphase}, or do not correspond to solutions of the Friedmann equations~\eqref{eqn:friedmann} in terms of the original variables \((H, \rho_m, \rho_r)\), since they are obtained from a singular coordinate transformation. Here we find the following conditions:
\begin{stmt}\label{stmt:fixsing}
A point \((H^{\star}, X^{\star})\) is an irregular fixed point of the dynamical system, if it satisfies one of the following criteria:
\begin{enumerate}[i.]
\item\label{it:fpsingisoat}
In the case \(X^{\star} = 1\), \(H^{\star} < 0\), \(W^{\star} \to \infty\) it is an isolated attractor. Trajectories approaching this point undergo an exponential decreasing of the scale factor~\eqref{eqn;expscalef}, while the matter density and pressure grow exponentially.
\item\label{it:fpsingisore}
In the case \(X^{\star} = 1\), \(H^{\star} > 0\), \(W^{\star} \to \infty\) it is an isolated repeller. Trajectories originating from this point undergo an exponential growth of the scale factor~\eqref{eqn;expscalef}, while the matter density and pressure decrease exponentially.
\item\label{it:fpsingisosa}
In the case \(X^{\star} = 0\), \(H^{\star} \neq 0\), \(W^{\star} \to \infty\) it is an isolated saddle point. Note that this point is neither a physical solution, nor approached by any trajectories.
\item\label{it:fpsingindet}
Fixed points, whose stability cannot be determined from a linearised analysis, are given by:
\begin{enumerate}[a.]
\item\label{it:fpsingidnz}
\(X^{\star} \in \{0, 1\}\), \(H^{\star} \neq 0\), \(W^{\star} > 0\) and \(W_H^{\star}\) diverges; even though this point lies inside the physical phase space spanned by the variables \((H, X)\), it does not have a corresponding solution in the original matter variables \((H, \rho_m, \rho_r)\), as it originates from a singular coordinate transformation.
\item\label{it:fpsingidwd}
\(H^{\star} = 0\) and \(W^{\star} \to \infty\); this point does not belong to the physical phase space and corresponds to a static universe with an infinite matter density.
\end{enumerate}
\end{enumerate}
\end{stmt}
An overview of all conditions listed in statements~\ref{stmt:fixreg} and~\ref{stmt:fixsing}, ordered by the properties of points in the compactified phase space, is given in figure~\ref{fig:index} in appendix~\ref{app:index}.

\subsection{Possibility of bounce and turnaround}\label{ssec:bounce}
We now come to the discussion of bounces and turnarounds, i.e., transitions between expanding and contracting phases of the evolution of the universe. Note that for any such transition we have \(\dot{a} = 0\), and hence \(H = 0\). Thus, these kind of transitions can occur only if \(H = 0\) lies inside the physical phase space~\eqref{eqn:phaseh} given by the condition that the total matter energy density is positive, and further require \(\dot{H} \neq 0\). The former is the case if and only if
\begin{equation}
W|_{H = 0} \geq 0\,.
\end{equation}
A bounce is given when \(\dot{H}|_{H = 0} > 0\), while a turnaround is characterized by \(\dot{H}|_{H = 0} < 0\). From the dynamics~\eqref{eqn:hdynh} of the Hubble parameter follows that \(\dot{H}\) is nonzero and finite at \(H = 0\) if and only if
\begin{equation}
\lim_{H \to 0}\frac{H}{(\ln W)_H} = \left.\frac{1}{(\ln W)_{HH}}\right|_{H = 0}
\end{equation}
is finite, and hence in particular \((\ln W)_H \to 0\) for \(H \to 0\). Explicitly calculating the derivatives of \(\ln W\) then shows that \(\dot{H} \neq 0\) if and only if \(W > 0\), \(W_H = 0\) and \(W_{HH} \neq 0\) at \(H = 0\), and that the sign of \(W_{HH}\) determines the sign of \(\dot{H}\). We thus conclude and summarize:

\begin{stmt}\label{stmt:bounce}
At \(H = 0\) we have \(\dot{H} \neq 0\) if and only if \(W > 0\), \(W_H = 0\) and \(W_{HH} \neq 0\), where
\begin{enumerate}[i.]
\item\label{it:bounce}
for \(W_{HH} < 0\) we have \(\dot{H} > 0\) and hence a bounce,
\item\label{it:turnaround}
for \(W_{HH} > 0\) we have \(\dot{H} < 0\) and hence a turnaround.
\end{enumerate}
\end{stmt}

Examples of bouncing cosmologies in $f(T)$ gravity have been discussed in Refs.~\cite{Cai:2011tc,Odintsov:2015uca,Bamba:2016gbu}.

\subsection{Impossibility of cyclic and oscillating universes}\label{ssec:cyclic}
Another interesting aspect, which is closely related to the existence of bounces and turnarounds as discussed in the previous section, is the possibility of cyclic universe solutions. Conventionally, these are defined as periodic solutions for the scale factor \(a(t + t_0) = a(t)\), and thus in particular imply that also the Hubble parameter \(H(t + t_0) = H(t)\) is periodic and has both positive and negative phases during each period. This means that in a cyclic universe both bounces and turnarounds occur periodically. However, this can immediately be excluded using statement~\ref{stmt:bounce}, since the conditions for a bounce and a turnaround are mutually exclusive and cannot be simultaneously satisfied for any given \(f(T)\) theory of gravity. Note that this property is even more restrictive and prohibits any solutions in which the scale factor shows an oscillating behavior in the sense that the dynamics change more than once between expansion and contraction.

We can also relax the periodicity condition and demand only that \(H(t + t_0) = H(t)\) is periodic. This allows for a periodic growth of the scale factor, \(a(t + t_0) = \lambda a(t)\) with constant \(\lambda\). However, one easily sees that also this is not possible. Recall from section~\ref{ssec:phase} that the sign of \(\dot{H}\) given by equation~\eqref{eqn:hdynh} is independent of \(X \in [0, 1]\). Any line of constant \(H\) can therefore be crossed in only one direction, either with increasing or decreasing \(H\), but not in both directions, as it would be necessary for a periodic orbit with variable \(H\). Periodic orbits with constant \(H\) and only variable \(X\) are likewise excluded, since for any fixed \(H\) also the sign of \(\dot{X}\) is independent of \(X\), and the same argument holds. Finally, oscillating behavior of the Hubble parameter \(H\) is also excluded, which follows from the same argumentation as for excluding the oscillating behavior of \(a\). We summarize:

\begin{stmt}
Periodic and oscillating orbits in the \((H, X)\) phase space, as well as cyclic and oscillating universe solutions are not possible.
\end{stmt}

We finally remark that this very general result does not depend in any way on the choice of the function \(f(T)\) in the action. It does, however, depend on the matter content, which we have fixed to dust and radiation. Exotic matter, which would allow for transitions between positive and negative matter densities, could potentially lead to oscillating behavior. However, we will not consider exotic matter here, and conclude our discussion of the phase space of \(f(T)\) gravity and its basic properties. Another important aspect is the existence and classification of finite time singularities. We present an exhaustive treatment in the following section.

\section{Finite time singularities}\label{sec:singular}
The dynamical systems approach detailed in the previous section now allows us to discuss the possibility of finite time singularities~\cite{Nojiri:2005sx,Nojiri:2008fk} in \(f(T)\) gravity. Note that there are different types of singularities, which can be distinguished by the behavior of \(H\) and \(\dot{H}\) near the singularity. This will be explained in detail in section~\ref{ssec:singtypes}. We then describe three types of singularities: those for which both \(H\) and \(\dot{H}\) become infinite are discussed in section~\ref{ssec:singii}. The case that \(\dot{H}\) diverges at a finite value of \(H\) is studied in section~\ref{ssec:singfi}. Finally, in section~\ref{ssec:singff} we consider the case that both \(H\) and \(\dot{H}\) remain finite, but higher time derivatives of \(H\) diverge.

\subsection{Types of singularities}\label{ssec:singtypes}
We start with a brief review of the possible types of finite time singularities, studied in detail in Refs.~\cite{Bamba:2012vg,ElHanafy:2017xsm}. For a singularity occurring at time \(t^{\diamond}\), it is conventional to approximate the Hubble parameter close to the singularity by the asymptotic behavior~\cite{Nojiri:2008fk}
\begin{equation}\label{eqn:singasymp}
H(t) \approx H^{\diamond} + \frac{h}{|t - t^{\diamond}|^k}
\end{equation}
with real constants \(H^{\diamond}, h, k\). Different types of singularities are distinguished by the value of the parameter \(k\). Classically one considers four types of singularities, which are denoted as follows~\cite{Nojiri:2005sx}:
\begin{enumerate}[Type I:]
\item\label{it:singtype1}
For \(k \geq 1\) both \(H\) and \(\dot{H}\) diverge for \(t \to t^{\diamond}\), so that in this case we can set \(H^{\diamond} = 0\) without loss of generality. By integrating the relation~\eqref{eqn:singasymp} one can see that also the logarithm of the scale factor \(\ln a\) diverges at the singularity, and so either \(a \to 0\) or \(a \to \infty\). If this singularity occurs in the past of an expanding universe, it is called a Big Bang. A future expanding singularity of this type is known as a Big Rip, while a future collapsing singularity is called a Big Crunch.

\item\label{it:singtype2}
In the range \(-1 < k < 0\) the Hubble parameter \(H \to H^{\diamond}\) stays finite, but its derivative \(\dot{H}\) diverges at the singularity. These singularities are called sudden singularities.

\item\label{it:singtype3}
If the singularity parameter is in the interval \(0 < k < 1\), we have a similar behavior to the case of a type~\ref{it:singtype1} singularity, so that both \(H\) and \(\dot{H}\) diverge. The only difference between these two types lies in the fact that for a singularity of type~\ref{it:singtype3} the scale factor remains finite.

\item\label{it:singtype4}
Finally, for \(k < -1\) with \(k \notin \mathbb{Z}\) both \(H\) and \(\dot{H}\) remain finite at the singularity, but higher time derivatives of \(H\) diverge.
\end{enumerate}
In principle it is also possible to consider singularities with a more general asymptotic behavior of the Hubble parameter than the power law~\eqref{eqn:singasymp}; however, we do not consider such general singularities here, and restrict ourselves to the four aforementioned types. Moreover, since singularities of type~\ref{it:singtype1} and type~\ref{it:singtype3} differ only by the asymptotic behavior of the scale factor \(a\), which is not explicit in our dynamical system, we will treat them together. By solving the asymptotic behavior~\eqref{eqn:singasymp} for the time \(t\) and doing the same with its time derivative, we can express \(\dot{H}\) through \(H\) in the vicinity of the singularity. Note that by definition of the constants we have
\begin{equation}\label{eqn:singposcond}
\frac{H - H^{\diamond}}{h} > 0\,,
\end{equation}
and so we can write
\begin{equation}\label{eqn:asympdh}
\dot{H} \approx \pm kh\left(\frac{H - H^{\diamond}}{h}\right)^{1 + \frac{1}{k}}\,,
\end{equation}
where the positive sign holds for future singularities \(t < t^{\diamond}\), while the negative sign holds for past singularities \(t > t^{\diamond}\). In the following, we use the abbreviation \(W^{\diamond} = W(H^{\diamond})\), as well as similar abbreviations for the derivatives of \(W\) at the singularity. We do not a priori demand that these derivatives exist at the singularity itself, but only in a neighborhood of the singularity, and then derive suitable limit values. In the following sections we give a detailed discussion of all singularity conditions. All conditions are also summarized in figure~\ref{fig:index} in appendix~\ref{app:index} in graphical form.

\subsection{Singularities of type I and III: $H \to \pm\infty$ and $\dot{H} \to \pm\infty$}\label{ssec:singii}
The first case we discuss is \(k > 0\), where both \(H\) and \(\dot{H}\) diverge at the singularity, and we set \(H^{\diamond} = 0\). From the asymptotic behavior~\eqref{eqn:asympdh} follows that the condition for a singularity can be expressed as
\begin{equation}\label{eqn:asympii}
0 = \pm\frac{1}{k}\lim_{H \to \pm\infty}\left(\frac{h}{H}\right)^{\frac{1}{k}} = \lim_{H = \pm\infty}\frac{H}{\dot{H}} = -\lim_{H \to \pm\infty}\frac{W_H}{(X + 3)W} = -\lim_{H \to \pm\infty}\frac{(\ln W)_H}{(X + 3)}\,,
\end{equation}
where the sign under the limit depends on whether one discusses a singularity for an expanding universe, \(H \to \infty\), or collapsing universe \(H \to -\infty\). We further see that also the type of the singularity, which is determined by the value of \(k\), can be read off from the asymptotic behavior of \((\ln W)_H\), by taking the logarithm under the limit~\eqref{eqn:asympii}. This can be summarized as follows:
\begin{stmt}\label{stmt:singii}
A finite time singularity with \(H \to \pm\infty\) and \(\dot{H} \to \pm\infty\) exists if and only if
\begin{equation}
\lim_{H \to \pm\infty}(\ln W)_H = 0\,,
\end{equation}
where the positive sign corresponds to an expanding universe, while the negative sign corresponds to a collapsing universe. The singularity parameter \(k > 0\) is given by
\begin{equation}
k = -\lim_{H \to \pm\infty}\frac{\ln|H|}{\ln|(\ln W)_H|}\,,
\end{equation}
with the same sign as above. The singularity lies in the past if asymptotically \(W_H > 0\), so that \(\sgn\dot{H} = -\sgn H\), and in the future if asymptotically \(W_H < 0\), so that \(\sgn\dot{H} = \sgn H\).
\end{stmt}

\subsection{Singularities of type II: finite $H$, but $\dot{H} \to \pm\infty$}\label{ssec:singfi}
We then discuss the case \(-1 < k < 0\), which is also called a sudden singularity, and which occurs when \(\dot{H}\) diverges for finite \(H = H^{\diamond}\). This case occurs when there exists \(H^{\diamond}\) such that
\begin{equation}\label{eqn:invdh}
0 = \lim_{H \to H^{\diamond}}\frac{1}{\dot{H}} = -\lim_{H \to H^{\diamond}}\frac{W_H}{(X + 3)HW}\,.
\end{equation}
Recall from the discussion of fixed points in section~\ref{ssec:fix} that \(\dot{H} \to 0\) whenever \(W \to 0\) or \(W \to \infty\). Hence, we can exclude these cases here and study only the case of a finite limit \(W \to W^{\diamond} > 0\). We distinguish the following two cases:

\begin{itemize}
\item
\(H^{\diamond} \neq 0\): In order for \(\dot{H}\) to become singular, the numerator \(W_H\) of~\eqref{eqn:asympdh} must vanish for \(H = H^{\diamond}\). This is the case if and only if \(W_H \to W_H^{\diamond} = 0\).

\item
\(H^{\diamond} = 0\): The case is similar to the aforementioned one, but the condition for a sudden singularity is more restrictive and reads
\begin{equation}
\lim_{H \to 0}\frac{W_H}{H} \to 0\,,
\end{equation}
which is the case if and only if both \(W_H\) and \(W_{HH}\) vanish at \(H = 0\).
\end{itemize}

In order to determine the singularity parameter \(k\), we make use of these conditions, which allow us to approximate the Friedmann function \(W\) near the singularity as
\begin{equation}\label{eqn:finsingpow}
W \approx W^{\diamond} + \epsilon\left(\frac{H - H^{\diamond}}{c}\right)^b\,,
\end{equation}
where \(\sgn c = \sgn h\) is chosen so that the expression inside the brackets becomes positive, \(\epsilon = \pm 1\) is a sign and we require \(b > 1\) for \(H^{\diamond} \neq 0\) and \(b > 2\) for \(H^{\diamond} = 0\). In this approximation the time derivative of the Hubble parameter becomes
\begin{equation}
\dot{H} \approx -\epsilon(X + 3)HW^{\diamond}\frac{c}{b}\left(\frac{H - H^{\diamond}}{c}\right)^{1 - b}\,.
\end{equation}

We now have to distinguish two different cases. For \(H^{\diamond} \neq 0\) we find that near the singularity \((H^{\diamond}, X^{\diamond})\) we have
\begin{equation}
\dot{H} \approx -\epsilon(X^{\diamond} + 3)H^{\diamond}W^{\diamond}\frac{c}{b}\left(\frac{h}{c}\right)^{1 - b}\left(\frac{H - H^{\diamond}}{h}\right)^{1 - b}\,.
\end{equation}
By comparison with the general form~\eqref{eqn:asympdh} we immediately read off \(k = -b^{-1}\) from the exponent. The sign in equation~\eqref{eqn:asympdh} can be read off from
\begin{equation}\label{eqn:singsignnz}
\sgn\left[-\epsilon\frac{(X^{\diamond} + 3)H^{\diamond}W^{\diamond}c}{khb}\left(\frac{h}{c}\right)^{1 - b}\right] = \sgn(\epsilon H^{\diamond}) = \begin{cases}
1 & \text{for future singularities,}\\
-1 & \text{for past singularities,}
\end{cases}
\end{equation}
where we have simply left out any positive, constant factors.

For \(H^{\diamond} = 0\) we find the approximation
\begin{equation}
\dot{H} \approx -\epsilon(X^{\diamond} + 3)W^{\diamond}\frac{c^2}{b}\left(\frac{h}{c}\right)^{2 - b}\left(\frac{H}{h}\right)^{2 - b}\,.
\end{equation}
In this case we read off \(k = (1 - b)^{-1}\), and the sign in equation~\eqref{eqn:asympdh} is given by
\begin{equation}\label{eqn:singsignz}
\sgn\left[-\epsilon\frac{(X^{\diamond} + 3)W^{\diamond}c^2}{khb}\left(\frac{h}{c}\right)^{2 - b}\right] = \sgn(\epsilon H) = \begin{cases}
1 & \text{for future singularities,}\\
-1 & \text{for past singularities,}
\end{cases}
\end{equation}
where in addition we used \(\sgn h = \sgn H\) in this case.

We finally remark on the sign \(\epsilon\) which appears in the results~\eqref{eqn:singsignnz} and~\eqref{eqn:singsignz}. Since \(W_H^{\diamond} = 0\), the Friedmann function \(W\) must have either an extremal point or an inflection point at \(H = H^{\diamond}\). In case of a maximum (minimum), \(\epsilon\) is positive (negative) on both sides of the singularity. If \(W\) has an inflection point, \(\epsilon\) differs on both sides of the singularity. We conclude and summarize:

\begin{stmt}\label{stmt:singfi}
A sudden singularity occurs at \(H = H^{\diamond}\) if and only if \(W^{\diamond} > 0\), \(W_H^{\diamond} = 0\) and
\begin{enumerate}[i.]
\item\label{it:singfinz}
either \(H^{\diamond} \neq 0\), in which case the singularity parameter is \(k = -b^{-1}\) with \(b > 1\),
\item\label{it:singfiz}
or \(H^{\diamond} = 0\) and \(W_{HH}^{\diamond} = 0\), in which case \(k = (1 - b)^{-1}\) with \(b > 2\),
\end{enumerate}
where \(b\) can be determined from the ansatz~\eqref{eqn:finsingpow}. The singularity occurs in the future in the following cases:
\begin{enumerate}[a.]
\item\label{it:singminpos}
\(W\) has a local minimum at \(H^{\diamond} > 0\),
\item\label{it:singmaxneg}
\(W\) has a local maximum at \(H^{\diamond} < 0\),
\item\label{it:singrise}
\(W\) has a rising inflection point at \(H^{\diamond}\) and the singularity is approached from \(|H| > |H^{\diamond}|\),
\item\label{it:singfall}
\(W\) has a falling inflection point at \(H^{\diamond}\) and the singularity is approached from \(|H| < |H^{\diamond}|\).
\end{enumerate}
The singularity occurs in the past if any of the aforementioned conditions is satisfied for \(-W\) instead of \(W\).
\end{stmt}

\subsection{Singularities of type IV: finite $H$ and $\dot{H}$}\label{ssec:singff}
We finally come to the case \(k < -1\) with \(k \notin \mathbb{Z}\), which is the most subtle type of singularity, since both \(H\) and \(\dot{H}\) remain finite, while higher time derivatives of \(H\) diverge. In order to study these singularities, we approximate the Hubble parameter as
\begin{equation}
H = H^{\diamond} + he^z
\end{equation}
near the singularity, where we made use of the positivity condition~\eqref{eqn:singposcond}. We then find that near the singularity
\begin{equation}
\frac{d}{dz}\ln|\dot{H}| \approx \frac{d}{dz}\ln\left|kh\left(e^z\right)^{1 + \frac{1}{k}}\right| = 1 + \frac{1}{k}\,.
\end{equation}
The dynamical equation~\eqref{eqn:hdynh} for \(\dot{H}\) yields
\begin{equation}
\frac{d}{dz}\ln|\dot{H}| = he^z\left(\frac{1}{H^{\diamond} + he^z} - \left[\ln\left|(\ln W)_H\right|\right]_H\right) = (H - H^{\diamond})\left(\frac{1}{H} - \left[\ln\left|(\ln W)_H\right|\right]_H\right)\,.
\end{equation}
The asymptotic behavior is obtained by approaching the singularity \(H \to H^{\diamond}\), so that we conclude
\begin{equation}
1 + \frac{1}{k} = \lim_{H \to H^{\diamond}}(H - H^{\diamond})\left(\frac{1}{H} - \left[\ln\left|(\ln W)_H\right|\right]_H\right)\,.
\end{equation}
For the values of \(k\) we consider in this section, the limit must be an element of the set
\begin{equation}
(0, 1) \setminus \left\{\frac{1}{2}, \frac{2}{3}, \frac{3}{4}, \ldots\right\}\,.
\end{equation}
It is helpful to distinguish two different cases:

\begin{itemize}
\item
For \(H^{\diamond} \neq 0\), the term \(1/H\) does not contribute to the limit. In this case the singularity condition reads
\begin{equation}
\lim_{H \to H^{\diamond}}(H - H^{\diamond})\left[\ln\left|(\ln W)_H\right|\right]_H \in (-1, 0) \setminus \left\{-\frac{1}{2}, -\frac{2}{3}, -\frac{3}{4}, \ldots\right\}\,.
\end{equation}

\item
For \(H^{\diamond} = 0\), the contribution of the term \(1/H\) must be taken into account, so that the singularity condition becomes
\begin{equation}
\lim_{H \to H^{\diamond}}(H - H^{\diamond})\left[\ln\left|(\ln W)_H\right|\right]_H \in (0, 1) \setminus \left\{\frac{1}{2}, \frac{1}{3}, \frac{1}{4}, \ldots\right\}\,.
\end{equation}
\end{itemize}

Note that in order to satisfy these conditions, it is in particular necessary that the limit is finite and nonzero, which requires the asymptotic behavior
\begin{equation}
\left[\ln\left|(\ln W)_H\right|\right]_H = \frac{W_{HH}}{W_H} - \frac{W_H}{W} \sim \frac{1}{H - H^{\diamond}}
\end{equation}
near the singularity. To achieve this behavior, we use the same approximation~\eqref{eqn:finsingpow} as in the previous case of a sudden singularity. For \(b > 0\) we obtain
\begin{equation}
\lim_{H \to H^{\diamond}}(H - H^{\diamond})\left[\ln\left|(\ln W)_H\right|\right]_H = -1 + \lim_{H \to H^{\diamond}}bW^{\diamond}\left[W^{\diamond} + \epsilon\left(\frac{H - H^{\diamond}}{c}\right)^b\right]^{-1} = b - 1\,.
\end{equation}
For \(H^{\diamond} \neq 0\) we thus require \(b \in (0,1) \setminus \left\{\frac{1}{2}, \frac{1}{3}, \frac{1}{4}, \ldots\right\}\), which implies that \(W_H\) diverges for \(H \to H^{\diamond}\). Similarly, for \(H^{\diamond} = 0\) we require \(b \in (1,2) \setminus \left\{\frac{3}{2}, \frac{4}{3}, \frac{5}{4}, \ldots\right\}\), so that \(W_H \to W_H^{\diamond} = 0\), while \(W_{HH}\) diverges for \(H \to H^{\diamond}\). This can be summarized as follows:

\begin{stmt}\label{stmt:singff}
Singularities with finite \(H\) and \(\dot{H}\) occur at \(H = H^{\diamond}\) if and only if \(W^{\diamond} > 0\) and
\begin{enumerate}[i.]
\item\label{it:singffnz}
either \(H^{\diamond} \neq 0\) and \(W_H\) diverges, in which case the singularity parameter is \(k = b^{-1}\),
\item\label{it:singffz}
or \(H^{\diamond} = 0\), \(W_H^{\diamond} = 0\) and \(W_{HH}\) diverges, in which case \(k = (1 - b)^{-1}\),
\end{enumerate}
where \(b\) is given by the asymptotic behavior~\eqref{eqn:finsingpow}, provided that \(k \neq \mathbb{Z}\). The conditions for future and past singularities are the same as in statement~\ref{stmt:singfi}.
\end{stmt}

We finally remark that the conditions for the existence of this type of singularity in particular imply the existence of fixed points, since they satisfy the condition~\ref{it:fpregindet} in statement~\ref{stmt:fixreg}. These singularities hence comprise a special class of fixed points which are reached in finite time.

This concludes our discussion of finite time singularities. In the next section we will shift our focus to another aspect of \(f(T)\) cosmology and derive a number of observable parameters.

\section{Observational properties}\label{sec:observ}
In the final section about the general dynamical system approach we discuss how to relate the dynamical system to physical properties and observables of the cosmological model. For this purpose we study in particular two properties, namely the accelerating expansion of the universe in section~\ref{ssec:accel} and the barotropic index of an equivalent dark energy model and the possibility of crossing the phantom divide in section~\ref{ssec:efflambda}. Specific phases of accelerating expansion, in particular inflation and the observed late time acceleration, are discussed in section~\ref{ssec:inflation}. We finally show how several observational parameters, such as the Hubble parameter, deceleration parameter and density parameters, can be read off from the dynamical system, and further be used to constrain the Friedmann function \(W(H)\) and select a particular phase space trajectory in section~\ref{ssec:cosmopar}.

\subsection{Accelerating expansion}\label{ssec:accel}
An important question about any $f(T)$ gravity model is whether it supports an epoch of accelerated expansion of the universe, and whether there are transitions between deceleration and acceleration. From the definition of the Hubble parameter immediately follows that
\begin{equation}
\dot{H} = \frac{d}{dt}\frac{\dot{a}}{a} = \frac{\ddot{a}a - \dot{a}^2}{a^2} = \frac{\ddot{a}}{a} - H^2\,,
\end{equation}
and hence the acceleration is given by
\begin{equation}\label{eqn:accel}
\frac{\ddot{a}}{a} = H^2 + \dot{H} = H\left(H - (X + 3)\frac{W}{W_H}\right)\,.
\end{equation}
We now focus on the transition between acceleration and deceleration, and thus in particular on phase space trajectories passing through the line where \(\ddot{a} = 0\). Note that for \(H = 0\) this condition implies \(\dot{H} = 0\), and hence corresponds to a fixed point, so that there is no transition in this case. We thus find that transitions can occur only for \(H \neq 0\) with
\begin{equation}\label{eqn:acctrans}
H = (X + 3)\frac{W}{W_H} = \frac{X + 3}{(\ln W)_H}\,.
\end{equation}
One easily checks that this is possible only if \(W\) and \(W_H\) are finite and nonzero. To determine the direction of the transition, we further calculate the third derivative
\begin{equation}
2H\dot{H} + \ddot{H} = \frac{d}{dt}\frac{\ddot{a}}{a} = \frac{\dddot{a}a - \ddot{a}\dot{a}}{a^2} = \frac{\dddot{a}}{a} - H^3 - H\dot{H}\,.
\end{equation}
We only need to study \(\dddot{a}\) in the particular case \(\ddot{a} = 0\) given by the relation~\eqref{eqn:acctrans}. In this case the third derivative of the scale factor is given by
\begin{equation}\label{eqn:acctrsgn}
\left.\frac{\dddot{a}}{a}\right|_{\ddot{a} = 0} = \frac{6H^3(X + 1)W - H^5W_{HH}}{(X + 3)W}\,.
\end{equation}
We can thus summarize:

\begin{stmt}\label{stmt:accel}
Transitions between acceleration and deceleration can occur only at phase space points satisfying \(H(\ln W)_H = X + 3\), following from equation~\eqref{eqn:acctrans}, and the direction of the transition is determined by
\begin{equation}
\sgn\dddot{a} = \sgn\left\{H[6(X + 1)W - H^2W_{HH}]\right\}\,,
\end{equation}
following from equation~\eqref{eqn:acctrsgn}.
\end{stmt}

Studies of whether and how different models can incorporate accelerated expansion include Refs.~\cite{Bengochea:2008gz,Linder:2010py,Wu:2010xk,Wu:2010av,Bamba:2010wb,Yang:2010hw,Zhang:2011qp,Setare:2012ry,Qi:2014yxa,Paliathanasis:2016vsw}.

\subsection{Dark energy and the phantom divide}\label{ssec:efflambda}
If we compare the cosmological field equations~\eqref{eqn:friedmann} with the corresponding equations for a generic dark energy model in general relativity, which are given by
\begin{subequations}\label{eqn:decosmo}
\begin{align}
H^2 &= \frac{8\pi G}{3}(\rho_m + \rho_r + \rho_{DE})\,,\label{eqn:decons}\\
\dot{H} &= -4\pi G\left[\rho_m + \frac{4}{3}\rho_r + (1 + w_{DE})\rho_{DE}\right]\,,\label{eqn:dedyn}
\end{align}
\end{subequations}
then one can easily read off that \(f(T)\) gravity can be described as an effective dark energy model, where the effective energy density of dark energy is given by
\begin{equation}\label{eqn:dedensity}
\rho_{DE} = \frac{6H^2 - W}{16\pi G}\,,
\end{equation}
while its effective barotropic index takes the form
\begin{equation}\label{eqn:debarind}
w_{DE} = -1 - \frac{(X + 3)}{3}\left(1 - 12\frac{H}{W_H}\right)\left(1 - 6\frac{H^2}{W}\right)^{-1} = -1 - \frac{X + 3}{3}\frac{[\ln|W - 6H^2|]_H}{(\ln W)_H}\,.
\end{equation}
We are in particular interested in the question whether the barotropic index is smaller or larger than \(-1\), or whether there exists some \(H = H^{\times}\) where it changes dynamically between these two possibilities. The critical value \(w_{DE} = -1\) discriminates between so-called phantom and non-phantom dark energy, and is hence also known as the ``phantom divide''. For this purpose it is sufficient to study the sign, zeroes and poles of the second term. We can proceed similarly to the discussion of fixed points and singularities and distinguish a number of different cases. First, we consider \(H^{\times} \neq 0\), and study the following three particular cases for \(W^{\times}\):
\begin{itemize}
\item
If \(W\) diverges, then also \(W_H\) diverges. Both terms in brackets in equation~\eqref{eqn:debarind} approach \(1\), and the barotropic index approaches \(-2 - \frac{X}{3} \neq -1\).
\item
For \(W \to 0\) one can easily see from the first expression for \(w_{DE}\) in equation~\eqref{eqn:debarind} that \(w_{DE} \to -1\), unless \(W_H \to 0\) and the corresponding factor in brackets diverges. However, one can see from the last expression in equation~\eqref{eqn:debarind} that also in the case \(W \to 0\) and \(W_H \to 0\) the term \((\ln W)_H\) in the denominator diverges as discussed in section~\ref{ssec:fix}, while the numerator stays finite. Hence, also in this case \(w_{DE} \to -1\). However, recall from section~\ref{ssec:fix} that \(W = 0\) implies \(\dot{H} = 0\), so that no crossing can occur in this case.
\item
If \(W \to 6H^2\), it follows from an analogous argument that the denominator \([\ln|W - 6H^2|]_H\) diverges, and hence also \(w_{DE}\) diverges.
\end{itemize}
We thus see that none of these cases allow for a crossing of the phantom divide. For any other value of \(W\), which is not covered by the aforementioned special cases, the second bracketed term in equation~\eqref{eqn:debarind} is nonzero, finite and not equal \(1\). We then need to consult the value of \(W_H\). Also here there are three particular cases to be discussed:
\begin{itemize}
\item
For \(W_H \to 12H\) the first bracketed term in equation~\eqref{eqn:debarind} vanishes and we obtain \(w_{DE} = -1\). Since \(\dot{H} \neq 0\) in this case, as follows from statement~\ref{stmt:fixreg}, this allows for a crossing of the phantom divide.
\item
For \(W_H \to 0\) the first bracketed term in equation~\eqref{eqn:debarind} diverges, and so does \(w_{DE}\).
\item
Finally, if \(W_H\) diverges, the first bracketed term in equation~\eqref{eqn:debarind} approaches \(1\), and we find \(w_{DE} \neq -1\).
\end{itemize}
Note that for all other values of \(W_H\) we likewise find \(w_{DE} \neq -1\), so that the only case we found so far for crossing the phantom divide is the one involving \(W_H = 12H\).

We are left with the case \(H = 0\). Since we are interested in the possibility of crossing the phantom divide, we need to consider only such cases in which we obtain a finite, nonzero \(\dot{H}\), i.e., a bounce or turnaround. These cases follow from the conditions given in statement~\ref{stmt:bounce}. In particular, we must have \(W > 0\), so that the second bracketed term in equation~\eqref{eqn:debarind} always approaches \(1\) for \(H \to 0\). Thus, in order to cross the phantom divide, the first bracketed term must vanish. This is the case if and only if
\begin{equation}
\lim_{H \to 0}\frac{H}{W_H} = \frac{1}{W_{HH}|_{H = 0}} = \frac{1}{12}\,,
\end{equation}
i.e., if and only if \(W > 0\), \(W_H = 0\) and \(W_{HH} = 12\) at \(H = 0\).

We also determine in which direction the phantom divide is crossed. For this purpose we calculate the total time derivative \(\dot{w}_{DE}\). In the first crossing case, where \(H^{\times} \neq 0\), we find
\begin{equation}
\dot{w}_{DE} = -\frac{(X^{\times} + 3)^2(W_{HH}^{\times} - 12)(W^{\times})^2}{432H^{\times}[W^{\times} - 6(H^{\times})^2]}\,,
\end{equation}
while in the second crossing case, for \(H^{\times} = 0\), we obtain
\begin{equation}
\dot{w}_{DE} = -\frac{(X^{\times} + 3)^2W^{\times}W_{HHH}^{\times}}{864}\,.
\end{equation}
We can summarize our findings as follows:

\begin{stmt}\label{stmt:phcross}
Crossing of the phantom divide occurs at \(H = H^{\times}\) if and only if \(W^{\times} > 0\) and
\begin{enumerate}[i.]
\item\label{it:phcrnz}
either \(H^{\times} \neq 0\), \(W^{\times} \neq 6(H^{\times})^2\) and \(W_H^{\times} = 12H^{\times}\), in which case
\begin{equation}
\sgn\dot{w}_{DE} = -\sgn\frac{(W_{HH}^{\times} - 12)}{H^{\times}[W^{\times} - 6(H^{\times})^2]}\,,
\end{equation}
\item\label{it:phcrz}
or \(H^{\times} = 0\), \(W_H^{\times} = 0\) and \(W_{HH}^{\times} = 12\), in which case
\begin{equation}
\sgn\dot{w}_{DE} = -\sgn W_{HHH}^{\times}\,.
\end{equation}
\end{enumerate}
\end{stmt}

From the point of view of phantom dark energy and the divide line crossing different models were considered in Refs.~\cite{Yang:2010hw,Wu:2010av,Bamba:2010wb}.

\subsection{Inflation and late time acceleration}\label{ssec:inflation}
From our discussion of the effective dark energy content in the preceding section follows another interesting remark. We have seen that in the case \(W = 0\), which implies \(\dot{H} = 0\), we have \(w_{DE} = -1\). This leads to the following conclusion, using the effective dark energy density~\eqref{eqn:dedensity}:

\begin{stmt}\label{stmt:desitter}
At fixed points \((H^{\star}, X^{\star})\) with \(W^{\star} = 0\) the solution becomes a de Sitter vacuum solution, i.e., \(\rho_r = \rho_m = 0\) and \(w_{DE} = -1\), with cosmological constant
\begin{equation}
\Lambda = 8\pi G\rho_{DE} = 3(H^{\star})^2\,.
\end{equation}
\end{stmt}

If \(H > 0\), then this solution models the observed late time acceleration of the universe. Note that de Sitter fixed points of this type \(W^{\star} = 0\) cannot be used to model inflation without invoking further mechanisms beyond the \(f(T)\) dynamics we study here. To see this, recall from statement~\ref{stmt:fixreg} that for \(W^{\star} = 0\) at \(H^{\star} > 0\) there exists a saddle point at \(X^{\star} = 1\) and an attractor at \(X^{\star} = 0\). Any trajectories in the vicinity of these fixed points ultimately converge to the attractor, and thus never leave the accelerating de Sitter phase. Hence, there would be no exit from this type of inflation.

We finally remark that fixed points with \(W^{\star} \to \infty\) could be potential candidates to model inflation. In the case \(H^{\star} > 0\) we see from statement~\ref{stmt:fixsing} that there exists a repeller at \(X^{\star} = 1\). This point, which is the limiting point of trajectories in their infinite past, corresponds to an infinite matter density. In this limit the scale factor \(a\) asymptotically becomes \(0\), with asymptotically constant Hubble parameter \(H = H^{\star}\). However, note that of course our purely classical model breaks down as soon as densities become sufficiently high that the quantum nature of matter becomes relevant, so that one cannot extrapolate this trajectory into the infinite past. We would rather expect that inflation starts from a quantum regime.

\subsection{Cosmological parameters}\label{ssec:cosmopar}
A number of observable parameters can be derived directly from the equations constituting the dynamical system. Most important are the density parameters, which are defined with the help of the critical density
\begin{equation}\label{eqn:critdens}
\rho_c = \frac{3H^2}{8\pi G}\,.
\end{equation}
For different contributions to the matter density present in our model this yields the straightforward definitions
\begin{equation}\label{eqn:denspar}
\Omega_m = \frac{\rho_m}{\rho_c} = \frac{(1 - X)W}{6H^2}\,, \quad
\Omega_r = \frac{\rho_r}{\rho_c} = \frac{XW}{6H^2}\,, \quad
\Omega_{DE} = \frac{\rho_{DE}}{\rho_c} = 1 - \frac{W}{6H^2}\,.
\end{equation}
Conventionally, one also defines a parameter \(\Omega_k\) related to the spatial curvature; however, this parameter vanishes identically for our model, since we restrict ourselves to spatially flat FLRW spacetimes. As a consequence, the parameters satisfy the constraint equation
\begin{equation}
\Omega_m + \Omega_r + \Omega_{DE} = 1\,,
\end{equation}
which is simply a rewriting of the corresponding Friedmann equation~\eqref{eqn:decons}, and which is in good agreement with current observations~\cite{Ade:2015xua}.

Another important set of observable parameters are of course the Hubble parameter \(H\) itself and the deceleration parameter \(q\) defined by
\begin{equation}
q = -\frac{\ddot{a}}{aH^2} = -1 - \frac{\dot{H}}{H^2} = -1 + (X + 3)\frac{W}{HW_H}\,.
\end{equation}
Their present values are related to a Taylor expansion of the scale factor \(a(t)\) around the present time \(t_0\) given by
\begin{equation}
a(t) = a_0\left[1 + H_0(t - t_0) - \frac{1}{2}H_0^2q_0(t - t_0)^2\right] + \mathcal{O}\left((t - t_0)^3\right)\,.
\end{equation}

From the present time values of these parameters we can derive two types of constraints. The Hubble parameter \(H_0\) and the ratio between radiation and matter given by
\begin{equation}
\frac{\Omega_{r,0}}{\Omega_{m,0}} = \frac{X_0}{1 - X_0}
\end{equation}
fix a point \((H_0, X_0)\) in phase space, and hence a particular trajectory. The total matter density further determines the present time value of \(W\) via
\begin{equation}
W_0 = 6H_0^2(\Omega_{m,0} + \Omega_{r,0})\,,
\end{equation}
while the present time value of \(W_H\) follows from the deceleration parameter as
\begin{equation}
W_{H,0} = \frac{(X_0 + 3)W_0}{H_0(1 - q_0)}\,.
\end{equation}
Hence, we obtain the first two coefficients in the Taylor expansion of the Friedmann function \(W\) around the present time value \(H_0\) of the Hubble parameter.

With this we finish our discussion of observational properties, and thereby of generic \(f(T)\) theories. We have now constructed a comprehensive set of tools to analyze any given \(f(T)\) theory by the properties of the corresponding Friedmann function \(W(H)\). In order to demonstrate the potential of these tools, we apply our formalism to a particular class of theories, for which \(F(T)\) is given by a power law, in the next section.

\section{Example: Power law model $F(T) = \alpha(-T)^n$}\label{sec:example}
After discussing the general formalism we developed in the previous sections, we now apply this formalism to a specific model. In this model \(F(T)\) is given by a power law \(F(T) = \alpha(-T)^n\), where \(\alpha\) and \(n\) are constant parameters. The main purpose of this section is to illustrate our formalism. We start with the definition of the power law model and a discussion of the allowed ranges and special values of its parameters in section~\ref{ssec:exdefi}. We then derive a number of properties of the general power law model, without restrictions on the constant parameters, using our general formalism. We discuss the physical phase space, its boundaries and fixed points in section~\ref{ssec:exphas}. Finite time singularities are discussed in section~\ref{ssec:exsing}. We continue by discussing phases of accelerating expansion in section~\ref{ssec:exaccel} and the properties in section~\ref{ssec:exphantom}. The results obtained in sections~\ref{ssec:exphas} to~\ref{ssec:exphantom} will then allow us to discuss the general dynamics and draw qualitative phase diagrams in section~\ref{ssec:exdyn}, as well as study the behavior of physical quantities along trajectories in section~\ref{ssec:extrajec}. Finally, in section~\ref{ssec:exobs} we derive a number of observational parameters and discuss their values for commonly used values of the model parameters.

\subsection{Definition and model parameters}\label{ssec:exdefi}
We start with a discussion of the generic properties of the power law model \(F(T) = \alpha (-T)^n\), where $\alpha$ and $n$ are constant parameters, whose values we leave arbitrary in this section. Note that \(-T = 6H^2\) is never negative. We calculate the function
\begin{equation}\label{eqn:powerw}
W = 6H^2 + (1 - 2n)\alpha(6H^2)^n\,,
\end{equation}
its first derivative
\begin{equation}\label{eqn:powerwh}
W_H = 12H + 2n(1 - 2n)\alpha\frac{(6H^2)^n}{H}
\end{equation}
and second derivative
\begin{equation}\label{eqn:powerwhh}
W_{HH} = 12\left[1 - n(1 - 2n)^2\alpha(6H^2)^{n - 1}\right]\,.
\end{equation}
It is important to distinguish a few special cases:
\begin{itemize}
\item
For $\alpha = 0$ the model trivially reduces to teleparallel equivalent of general relativity.
\item
In the case $n = 0$ we see that $F(T)$ is simply a cosmological constant. Hence, the model becomes equivalent to GR with a cosmological constant.
\item
For $n = 1$ the total Lagrangian reads \((1 - \alpha)T\). This model is equivalent to general relativity, where the gravitational constant is rescaled by a factor \(1 - \alpha\). Obviously, \(\alpha = 1\) must be excluded in this case.
\item
For \(n = \frac{1}{2}\) the terms originating from \(F(T)\) do not contribute to the cosmological field equations. Hence, we obtain the same cosmological dynamics as for general relativity.
\end{itemize}
In the following we will exclude these special cases from our analysis, since they would have to be treated separately, but do not yield any new features beyond the exhaustively studied GR cosmology with a cosmological constant.

\subsection{Physical phase space and fixed points}\label{ssec:exphas}
We now apply the general framework we developed in this article to this model. We first determine the physical phase space, which is given by the condition \(W \geq 0\) obtained in section~\ref{ssec:phase}. The boundary of the phase space is thus given by the condition that either \(W = 0\) or \(W \to \infty\), which are also fixed point conditions according to statements~\ref{stmt:fixreg} and~\ref{stmt:fixsing} detailed in section~\ref{ssec:fix}. It thus makes sense first study the fixed point conditions. We start with the condition \(W = 0\), which for the model we consider here has the solutions
\begin{equation}\label{eqn:exfpcond}
H = 0 \quad \text{and} \quad (6H^2)^{1 - n} = (2n - 1)\alpha\,.
\end{equation}
The point \(H = 0\), which is a solution only for \(n > 0\), is special and will be treated separately. The second equation has a positive and a negative solution for \((2n - 1)\alpha > 0\) and \(n \neq 1\). In the following we will denote the positive solution by \(H = H^{\star}\). According to conditions~\ref{it:fpregisoat}, \ref{it:fpregisore} and~\ref{it:fpregisosa} in statement~\ref{stmt:fixreg} we thus obtain different fixed points \((H,X) \in \mathcal{P}\): one attractor \((H^{\star}, 0)\), one repeller \((-H^{\star}, 0)\) and two saddle points \((\pm H^{\star}, 1)\). According to statement~\ref{stmt:desitter} they all correspond to de Sitter universe. Note that these are the only fixed points with \(H \neq 0\), since the power law model does not satisfy any other condition listed in statement~\ref{stmt:fixreg} or~\ref{stmt:fixsing}, except for \(H = 0\). The point $(H^\star, 0)$ should correspond to the same state as the stable de Sitter fixed point found in studies with Hubble-rescaled variables~\cite{Wu:2010xk,Zhang:2011qp,Setare:2012ry,Feng:2014fsa,Mirza:2017vrk}.

In the case \(H = 0\) we see that condition~\ref{it:fpregidwz} in statement~\ref{stmt:fixreg} is satisfied for \(n > 0\), while condition~\ref{it:fpsingidwd} in statement~\ref{stmt:fixsing} is satisfied for \(n < 0\). The first one is a regular vacuum fixed point, while the latter belongs to a divergent matter density and hence lies on the boundary outside of the physical phase space. In both cases the stability cannot be determined from the linearised analysis. We will determine their stability from the phase diagrams in section~\ref{ssec:exdyn}.

From our analysis follows that the physical phase space is bounded in \(H\) if either \(\alpha > 0\) and \(n > \frac{1}{2}\) or \(\alpha < 0\) and \(n < \frac{1}{2}\). Studying the sign of \(W\) in these cases yields different restrictions on the physical phase space:
\begin{itemize}
\item
For \(\alpha > 0\) and \(n > 1\) we obtain \(W \geq 0\) only for \(|H| \leq H^{\star}\). The absolute value of the Hubble parameter thus has an upper bound.
\item
For \(\alpha > 0\) and \(\frac{1}{2} < n < 1\) the physical phase space is given by \(|H| \geq H^{\star}\), and so we receive a lower bound instead.
\item
For \(\alpha < 0\) and \(n < \frac{1}{2}\) the absolute value of the Hubble parameter likewise has a lower bound given by \(|H| \geq H^{\star}\).
\end{itemize}
Finally, for \((2n - 1)\alpha \leq 0\), the physical phase space covers all of \(H \in \mathbb{R}\).

\subsection{Finite time singularities}\label{ssec:exsing}
We now come to the discussion of finite time singularities in the power law model, for which we proceed in the same way as for the fixed points. We start with the case \(H \to \pm\infty\) as discussed in section~\ref{ssec:singii}, which belongs to the boundary of the physical phase space unless \(\alpha > 1\) and \(n > 1\). From statement~\ref{stmt:singii} follows that we need to consider the asymptotic behavior of \((\ln W)_H\) for \(H \to \pm\infty\). Note that in this limit we have \(W \sim H^2\) if \(n < 1\) and \(W \sim H^{2n}\) if \(n > 1\). In both cases, we find the asymptotic behavior
\begin{equation}
(\ln W)_H \sim \frac{1}{H}\,,
\end{equation}
which corresponds to a finite time singularity of type~\ref{it:singtype1} with parameter \(k = 1\). For \(H \to \infty\) we find \(W_H > 0\), so that this is a past singularity, and hence a Big Bang. Conversely, for \(H \to -\infty\), we find \(W_H < 0\) and this is a future singularity for a collapsing universe, hence a Big Crunch.

We then come to sudden singularities, or singularities of type~\ref{it:singtype2}, as discussed in section~\ref{ssec:singfi}, which occur at finite \(H\). Recall from statement~\ref{stmt:singfi} that these occur only where \(W^{\diamond}\) is nonzero and finite, so that we can exclude \(H = 0\) for the power law model from our discussion. We are thus looking for points \(H^{\diamond} \neq 0\) where \(W_H^{\diamond} = 0\), as also required by statement~\ref{stmt:singfi}. This condition yields
\begin{equation}\label{eqn:exsingcond}
(6H^2)^{1 - n} = n(2n - 1)\alpha\,,
\end{equation}
where the right hand side is nonzero since we have already excluded those values for \(n\) and \(\alpha\) for which it will vanish. In the following we will denote by \(H^{\diamond}\) the positive solution of this equation, if it exists. This is the case for the following parameter ranges:
\begin{itemize}
\item
For \(\alpha > 0\) and \(n < 0\) the physical phase space covers the whole range \(H \in \mathbb{R}\), and hence also contains the singularity \(H^{\diamond}\).
\item
For \(\alpha < 0\) and \(0 < n < \frac{1}{2}\) we find the singularity at \(H^{\diamond} < H^{\star}\); however, this point lies outside the physical phase space.
\item
For \(\alpha > 0\) and \(\frac{1}{2} < n < 1\) we have qualitatively the same situation as in the aforementioned case, with a singularity at \(H^{\diamond} < H^{\star}\) outside the physical phase space.
\item
For \(\alpha > 0\) and \(n > 1\) the singularity also satisfies \(H^{\diamond} < H^{\star}\), but in this case this point lies inside the physical phase space.
\end{itemize}
Hence, we need to discuss only the first and the last of these ranges. At the singularity we find that
\begin{equation}
W^{\diamond} = 6\left(1 - \frac{1}{n}\right)(H^{\diamond})^2\,, \quad W_H^{\diamond} = 0\,, \quad W_{HH}^{\diamond} = 24(1 - n)\,.
\end{equation}
In particular, we find that \(W_{HH}^{\diamond}\) is finite and nonzero, and so the asymptotic behavior of the Friedmann function \(W\) is given by
\begin{equation}
W - W^{\diamond} \sim (H - H^{\diamond})^2
\end{equation}
near the singularity. We hence obtain the singularity parameter \(k = -\frac{1}{2}\), which is in the expected range for a singularity of type~\ref{it:singtype2}. Finally, note that for \(n < 0\) we have \(W_{HH}^{\diamond} > 0\), so that \(W\) has a local minimum and we find a future sudden singularity at \(H^{\diamond} > 0\), which is complemented by a past sudden singularity at \(-H^{\diamond}\). The opposite time behavior is obtained in the case \(n > 1\).

\subsection{Accelerating expansion}\label{ssec:exaccel}
As the next aspect we discuss the possibility of an accelerating expansion and the transition between accelerating and decelerating phases, noted already in the early papers \cite{Bengochea:2008gz,Linder:2010py}. For the acceleration we find the expression
\begin{equation}\label{eqn:exaccel}
\frac{\ddot{a}}{a} = H^2\left[1 - \frac{X + 3}{2}\frac{6H^2 + (1 - 2n)\alpha(6H^2)^n}{6H^2 + n(1 - 2n)\alpha(6H^2)^n}\right]\,.
\end{equation}
To determine the behavior of this function on the physical phase space, it is helpful to introduce the auxiliary functions
\begin{equation}\label{eqn:uvdef}
V = \frac{HW_H}{2} = 6H^2 + n(1 - 2n)\alpha(6H^2)^n\,, \quad U = \frac{W}{V} = \frac{6H^2 + (1 - 2n)\alpha(6H^2)^n}{6H^2 + n(1 - 2n)\alpha(6H^2)^n}\,.
\end{equation}
With this definition it follows that
\begin{equation}
\frac{\ddot{a}}{a} = H^2\left[1 - \frac{X + 3}{2}U\right]\,,
\end{equation}
so that \(\ddot{a} > 0\) if and only if \(U < \frac{2}{X + 3}\), where $\frac{2}{X + 3}$ always takes values in the interval \([\frac{1}{2}, \frac{2}{3}]\). We have thus obtained a simple condition which determines the sign of the acceleration for the power law model. We will discuss this condition and its implications on the history of the universe in more detail in section~\ref{ssec:extrajec}.

\subsection{Dark energy and the phantom divide}\label{ssec:exphantom}
We start our discussion in this section with the possibility of crossing the phantom divide. For this purpose we check the conditions given in statement~\ref{stmt:phcross} given in section~\ref{ssec:efflambda}. Condition~\ref{it:phcrnz} is not satisfied, since there is no \(H^{\times} \neq 0\) for which \(W_H^{\times} = 12H^{\times}\), as follows from equation~\eqref{eqn:powerwh}. Also condition~\ref{it:phcrz} is not satisfied, since \(W_{HH} \neq 12\) at \(H = 0\), independent of the parameters of the power law model, as follows from equation~\eqref{eqn:powerwhh}. Hence, there is no crossing of the phantom divide, as noted before in Ref.~\cite{Wu:2010mn} via statefinder and \textit{Om} diagnostic.

The same result can also be seen from the barotropic index of the dark energy component, which is given by
\begin{equation}\label{eqn:exbarind}
w_{DE} = -1 + \frac{n}{3}(X + 3)\frac{6H^2 + (1 - 2n)\alpha(6H^2)^n}{6H^2 + n(1 - 2n)\alpha(6H^2)^n} = -1 + \frac{n}{3}(X + 3)U\,,
\end{equation}
where \(U\) is defined in equation~\eqref{eqn:uvdef}. Note that \(w_{DE} = -1\) if and only if \(W = 0\), and that this condition corresponds to fixed points according to our analysis in section~\ref{ssec:exphas}. Hence, there are no transitions between \(w_{DE} < -1\) and \(w_{DE} > -1\), and thus no crossing of the phantom divide.

Using the formula~\eqref{eqn:exbarind} we can also discriminate between phantom and non-phantom dark energy. One can easily see that \(w_{DE} < -1\) if and only if \(nU < 0\), while for \(nU > 0\) we find \(w_{DE} > -1\). If any of these conditions is satisfied for some point \((H, X) \in \mathcal{P}\), it is satisfied for all points on the trajectory through \((H, X)\), since there is no crossing of the phantom divide. We can thus distinguish between phantom and non-phantom trajectories. We will do so in detail in section~\ref{ssec:extrajec}

\subsection{General dynamics and phase diagrams}\label{ssec:exdyn}
We now use the results on the physical phase space and the existence and behavior of fixed points and singularities obtained in sections~\ref{ssec:exphas} and~\ref{ssec:exsing} in order to discuss the general dynamics of the cosmological model for different values of the parameters \(n\) and \(\alpha\). Note that the only values of \(H\) at which the sign of \(\dot{H}\) and \(\dot{X}\), and hence the qualitative behavior of the system can change, are the values \(0, \pm H^{\diamond}, \pm H^{\star}\). They divide the physical phase space into several regions, in which we now study the sign of \(\dot{H}\), as well as the aforementioned physical quantities. The qualitative phase diagrams derived from our analysis are shown in figure~\ref{fig:powerphasp}, where we have used gray lines in order to mark the following distinguished values of \(H\): a solid line marks \(H = 0\), dashed lines mark the singularities \(H = \pm H^{\diamond}\) and dotted lines mark the fixed points \(H = \pm H^{\star}\). Note that all diagrams are symmetric under the transformation \(H \mapsto -H, X \mapsto X, \dot{H} \mapsto \dot{H}, \dot{X} \mapsto -\dot{X}\). We will therefore only discuss the right half, \(H \geq 0\), which corresponds to an expanding phase of the universe. Then the diagrams can be classified as follows:

\begin{figure}[p]
\centering
\subfigure[$\alpha < 0$ and $n < \frac{1}{2}$]{\includegraphics[width=0.33\textwidth]{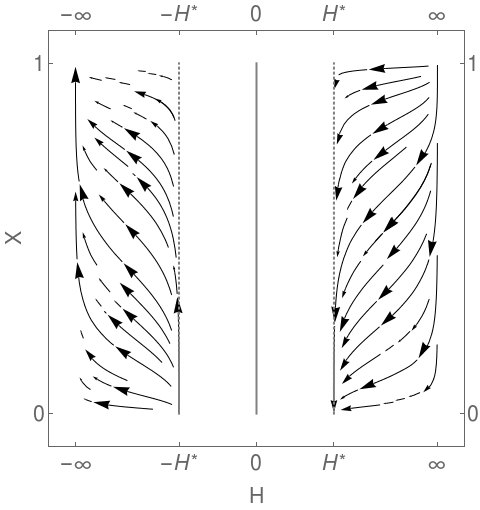}\label{fig:powerphasp1}}
\subfigure[$\alpha < 0$ and $n > \frac{1}{2}$]{\includegraphics[width=0.33\textwidth]{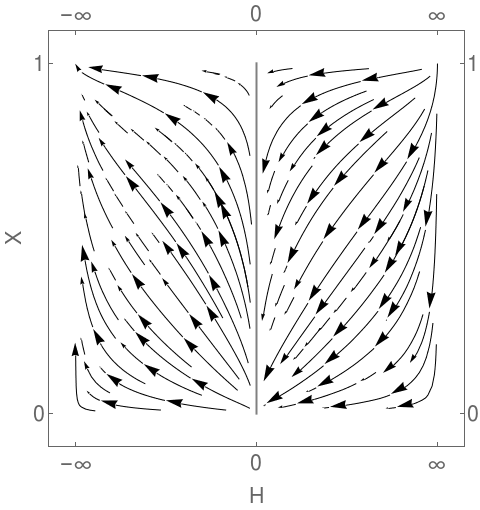}\label{fig:powerphasp2}}
\subfigure[$\alpha > 0$ and $n < 0$]{\includegraphics[width=0.33\textwidth]{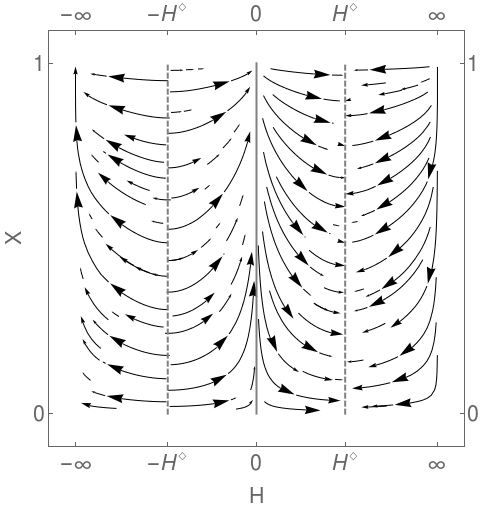}\label{fig:powerphasp3}}
\subfigure[$\alpha > 0$ and $0 < n < \frac{1}{2}$]{\includegraphics[width=0.33\textwidth]{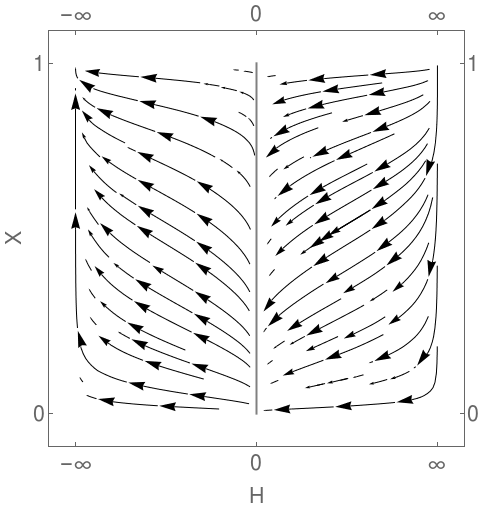}\label{fig:powerphasp4}}
\subfigure[$\alpha > 0$ and $\frac{1}{2} < n < 1$]{\includegraphics[width=0.33\textwidth]{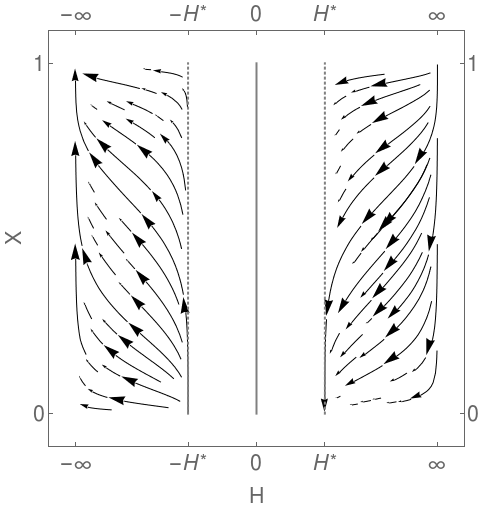}\label{fig:powerphasp5}}
\subfigure[$\alpha > 0$ and $n > 1$]{\includegraphics[width=0.33\textwidth]{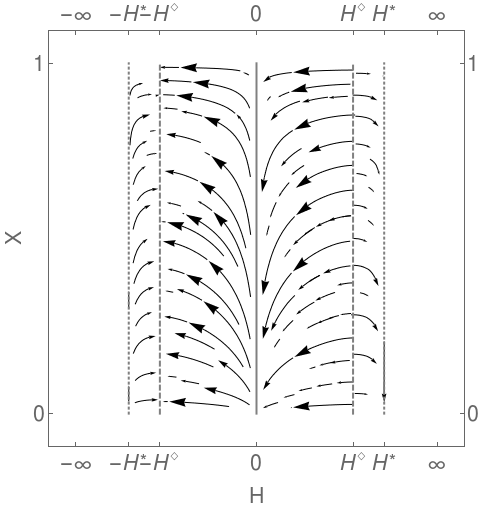}\label{fig:powerphasp6}}
\caption{Qualitative phase diagrams for the power law model. Distinguished values of \(H\) are marked by the following: a solid line marks \(H = 0\), dashed lines mark the singularities \(H = \pm H^{\diamond}\) and dotted lines mark the fixed points \(H = \pm H^{\star}\).}
\label{fig:powerphasp}
\end{figure}

\begin{itemize}
\item
For \(\alpha < 0\), \(n > \frac{1}{2}\) shown in figure~\ref{fig:powerphasp2} and \(\alpha > 0\), \(0 < n < \frac{1}{2}\) shown in figure~\ref{fig:powerphasp4} the phase diagrams are qualitatively identical. The region \(H > 0, 0 < X < 1\) is filled with trajectories which start from a Big Bang singularity at \((\infty, 1)\) and end at a static fixed point at \((0, 0)\). These are bounded by trajectories with \(X \equiv 0\) and \(X \equiv 1\), both of which start at the Big Bang \((\infty, X)\) and end at a static universe \((0, X)\). Finally, all points \((0, X)\) with \(X \in [0, 1]\) are fixed points. This fact does not immediately become apparent from the phase diagrams, since \(\dot{X} \sim H\) and \(\dot{H} \sim H^2\), so that \(\dot{H}/\dot{X} \sim H \to 0\) and trajectories become vertical near the fixed line \(H = 0\); however, the velocity with which these trajectories are traversed converges to 0.

\item
For \(\alpha < 0\), \(n < \frac{1}{2}\) shown in figure~\ref{fig:powerphasp1} and \(\alpha > 0\), \(\frac{1}{2} < n < 1\) shown in figure~\ref{fig:powerphasp5} there exists a critical value \(H = H^{\star}\). Physical trajectories in the region \(H > H^{\star}, 0 < X < 1\) start at the Big Bang singularity \((\infty, 1)\) and end at the attractive de Sitter fixed point \((H^{\star}, 0)\). Also in this case there exist bounding trajectories with \(X \equiv 0\) and \(X \equiv 1\) going from \((\infty, X)\) to \((H^{\star}, X)\). Finally, there exists another bounding trajectory connecting the de Sitter saddle point \((H^{\star}, 1)\) to the attractive de Sitter fixed point \((H^{\star}, 0)\).

\item
In the case \(\alpha > 0\), \(n < 0\) there exists another type of critical value \(H = H^{\diamond}\) corresponding to a sudden singularity, which splits the physical phase space into different parts. Trajectories in the region \(H > H^{\diamond}\) start at the Big Bang singularity \((\infty, 1)\) and reach the sudden singularity \((H^{\diamond}, X)\) at a finite value of \(X\). Points with \(0 < H < H^{\diamond}\) belong to trajectories starting at the static saddle point \((0, 1)\), which also reach the sudden singularity \((H^{\diamond}, X)\) at a finite value of \(X\).

\item
In the case \(\alpha > 0\), \(n > 1\) both types of critical values exist, with \(0 < H^{\diamond} < H^{\star}\). Trajectories in the region \(H^{\diamond} < H < H^{\star}\) start from the sudden singularity \((H^{\diamond}, X)\) at a finite value of \(X\) and approach the stable de Sitter fixed point \((H^{\star}, 0)\). In the region \(0 < H < H^{\diamond}\), trajectories have the same starting condition, but approach the static saddle point \((0, 0)\).
\end{itemize}

We can now also study the stability of the fixed points on the line \(H = 0\), which we identified in section~\ref{ssec:exphas}. Since the sign of \(\dot{H}\) is the same on both sides \(H > 0\) and \(H < 0\) of these fixed points, it follows that trajectories on one side are traversed towards \(H = 0\), while on the other side they are traversed away from \(H = 0\). Hence, the fixed points on the line \(H = 0\) are always saddle points.

\subsection{Physical trajectories and their properties}\label{ssec:extrajec}
Making use of the qualitative phase diagrams, we can now also study the behavior of the acceleration~\eqref{eqn:exaccel} and the barotropic index~\eqref{eqn:exbarind} of the effective dark energy along the physical trajectories. In this discussion we restrict ourselves to trajectories in the interior of the phase space and exclude those which are located entirely on the boundary. Also here we discuss only the case \(H \geq 0\), using the fact that \(\ddot{a}\) and \(w_{DE}\) do not change under a sign reversal \(H \mapsto -H\). For this purpose we now take a closer look at the function \(U(H)\) defined in equation~\eqref{eqn:uvdef}. Note that
\begin{equation}
V^2U_H = W_HV - WV_H = 12(n - 1)^2(2n - 1)\alpha H(6H^2)^n
\end{equation}
has no zeroes for \(H \neq 0\). Thus, \(U\) is monotonous whenever \(V\) is finite (and hence also \(U\) is defined). We distinguish the following cases, which are summarized in figure~\ref{fig:parranges}:
\begin{figure}[htb]
\begin{tikzpicture}[align=center,thick]
\draw (-7,-1.5) rectangle (7,7.5);
\draw (0,-1.5) to (0,7.5);
\draw (-7,0) to (7,0);
\draw (-7,1.5) to (7,1.5);
\draw (0,3) to (7,3);
\draw (-7,4.5) to (0,4.5);
\draw (-7,6) to (0,6);
\node at (-3.5,-0.75) {$(\infty, 1) \to (H^{\star}, 0)$: $\ddot{a}\nearrow$, $w_{DE} < -1$};
\node at (-3.5,0.75) {$(\infty, 1) \to (H^{\star}, 0)$: $\ddot{a}\nearrow$, $w_{DE} > -1$};
\node at (-3.5,3) {$(\infty, 1) \to (0, 0)$: $\ddot{a} < 0$, $w_{DE} > -1$};
\node at (-3.5,5.25) {$(\infty, 1) \to (0, 0)$: $\ddot{a} < 0$ or $\ddot{a}\nearrow\searrow$, $w_{DE} > -1$};
\node at (-3.5,6.75) {$(\infty, 1) \to (0, 0)$: $\ddot{a}\searrow$, $w_{DE} > -1$};
\node at (3.5,-0.75) {$(0, 1) \to (H^{\diamond}, X)$: $\ddot{a} > 0$, $w_{DE} > -1$;\\$(\infty, 1) \to (H^{\diamond}, X)$: $\ddot{a} < 0$, $w_{DE} < -1$};
\node at (3.5,0.75) {$(\infty, 1) \to (0, 0)$: $\ddot{a} < 0$, $w_{DE} > -1$};
\node at (3.5,2.25) {$(\infty, 1) \to (H^{\star}, 0)$: $\ddot{a}\nearrow$, $w_{DE} > -1$};
\node at (3.5,5.25) {$(H^{\diamond}, X) \to (0, 0)$: $\ddot{a} < 0$, $w_{DE} > -1$;\\$(H^{\diamond}, X) \to (H^{\star}, 0)$: $\ddot{a} > 0$, $w_{DE} < -1$};
\node at (-7,-2) {$-\infty$};
\node at (0,-2) {$0$};
\node at (7,-2) {$\infty$};
\node at (0,-2.5) {$\alpha$};
\node at (-7.5,-1.5) {$-\infty$};
\node at (-7.5,0) {$0$};
\node at (-7.5,1.5) {$\frac{1}{2}$};
\node at (-7.5,3) {$1$};
\node at (-7.5,4.5) {$\frac{3}{2}$};
\node at (-7.5,6) {$2$};
\node at (-7.5,7.5) {$\infty$};
\node at (-8,3) {$n$};
\end{tikzpicture}
\caption{Physical trajectories \((H_i, X_i) \to (H_f, X_f)\) in the power law model with \(H \geq 0\), classified by their asymptotic initial and final states \((H_i, X_i)\) and \((H_f, X_f)\). Due to the symmetry of the phase diagrams, each of these has a corresponding trajectory with \(H \leq 0\), which can be obtained by replacing \(H \mapsto -H\) and reversing the direction of the arrows. Here \(\ddot{a}\nearrow\) indicates a transition from deceleration to acceleration, while \(\ddot{a}\searrow\) indicates a transition in the opposite direction.}
\label{fig:parranges}
\end{figure}
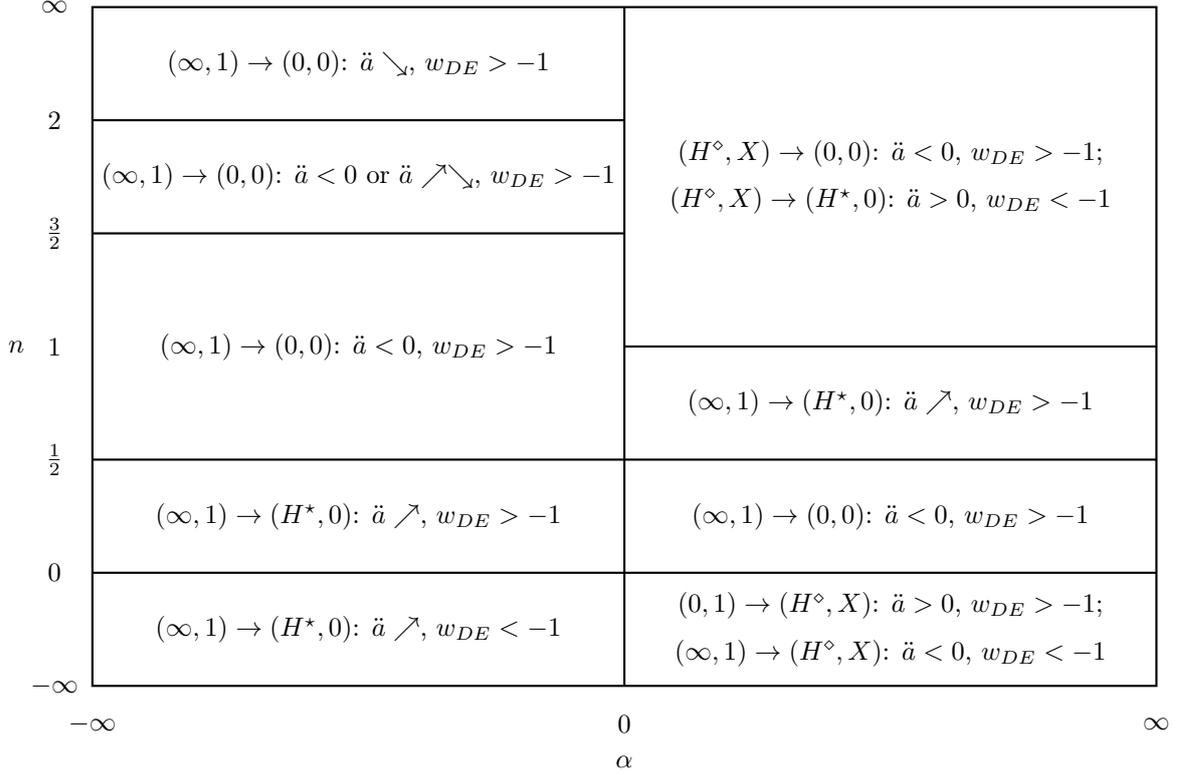
\begin{itemize}
\item
For \(\alpha < 0\) and \(n < \frac{1}{2}\) as shown in figure~\ref{fig:powerphasp1} we have \(U \to 0\) for the de Sitter fixed point \(H \to H^{\star}\), where \(\ddot{a} > 0\), and \(U \to 1\) for \(H \to \infty\), hence \(\ddot{a} < 0\). We thus have a transition from a decelerating to an accelerating phase. For \(n < 0\) we have phantom dark energy, \(w_{DE} < -1\), while for \(n > 0\) we obtain \(w_{DE} > -1\). Such dark energy behavior was noted in Ref.~\cite{Wu:2010mn}.

\item
For \(\alpha < 0\) and \(\frac{1}{2} < n < 1\) as shown in figure~\ref{fig:powerphasp2} we find \(U \to \frac{1}{n} > 1\) for \(H \to 0\) and \(U \to 1\) for \(H \to \infty\). Hence, we have \(1 < U < \frac{1}{n}\) everywhere and thus \(\ddot{a} < 0\). There is no accelerating phase. Further, \(w_{DE} > -1\), so that there is no phantom dark energy.

\item
For \(\alpha < 0\) and \(n > 1\), which is also shown in figure~\ref{fig:powerphasp2}, the limiting cases are given by \(U \to 1\) for \(H \to 0\) and \(U \to \frac{1}{n} < 1\) for \(H \to \infty\). The sign of the acceleration depends on the value of \(n\). For \(n < \frac{3}{2}\) follows that \(U > \frac{2}{3}\) everywhere, and hence \(\ddot{a} < 0\), so that there is no accelerating phase. For \(n > 2\) there is \(U < \frac{1}{2}\) for \(H \to \infty\), so that all trajectories pass from an accelerating to a decelerating phase. In the intermediate parameter range \(\frac{3}{2} \leq n \leq 2\) the accelerating phase does not cover the whole edge \(H \to \infty\) of the phase diagram, but only the part \(X < 2n - 3\). This means in particular that the Big Bang singularity at \(H \to \infty, X \to 1\) is located in the decelerating phase. Trajectories starting from the Big Bang singularity may either bypass the accelerating phase completely or experience a transient positive acceleration. In all of these cases we have \(w_{DE} > -1\).

\item
In the case \(\alpha > 0\) and \(n < 0\) shown in figure~\ref{fig:powerphasp3} we need to discuss two regions of the phase diagram separately. For \(H < H^{\diamond}\) there is \(V < 0\) and hence also \(U < 0\), so that \(\ddot{a} > 0\) and \(w_{DE} > -1\). For \(H > H^{\diamond}\), we find \(W > V > 0\), hence \(U > 1\), \(\ddot{a} < 0\) and \(w_{DE} < -1\). However, since these regions are separated by a singularity, there are no transitions between accelerating and decelerating phases or between phantom and non-phantom dark energy.

\item
For \(\alpha > 0\) and \(0 < n < \frac{1}{2}\) as shown in figure~\ref{fig:powerphasp4} we have the limiting cases \(U \to \frac{1}{n} > 2\) for \(H \to 0\) and \(U \to 1\) for \(H \to \infty\). We thus have \(U > 1\) everywhere and therefore \(\ddot{a} < 0\), so that there is no accelerating phase. There is also no phantom dark energy, since \(w_{DE} > -1\).

\item
When \(\alpha > 0\) and \(\frac{1}{2} < n < 1\) as shown in figure~\ref{fig:powerphasp5}, there is a de Sitter fixed point with \(U \to 0\) for \(H \to H^{\star}\), hence \(\ddot{a} > 0\), while \(U \to 1\) for \(H \to \infty\), and thus \(\ddot{a} < 0\). It follows that there is a transition from a decelerating to an accelerating phase. We still find \(w_{DE} > -1\) also in this case. Such dark energy behavior was noted in Ref.~\cite{Wu:2010mn}.

\item
Finally, in the case \(\alpha > 0\) and \(n > 1\) the physical phase space \(W \geq 0\) splits into two regions divided by a singularity at \(H = H^{\diamond}\), as shown in figure~\ref{fig:powerphasp6}, which we discuss separately. For \(H < H^{\diamond}\) we have \(W > V > 0\) and thus \(U > 1\), hence \(\ddot{a} < 0\) and \(w_{DE} > -1\). In contrast, for \(H > H^{\diamond}\) we find \(V < 0\), which yields \(U < 0\), thus \(\ddot{a} > 0\) and \(w_{DE} < -1\). Since accelerating and decelerating phases are separated by a singularity, there is no transition. The same holds for phantom and non-phantom dark energy.
\end{itemize}

\subsection{Observational properties}\label{ssec:exobs}
We finally discuss how to derive a number of observational parameters for the power law model. For the density parameters~\eqref{eqn:denspar} we find the expressions
\begin{subequations}
\begin{align}
\Omega_m &= (1 - X)\left[1 + (1 - 2n)\alpha(6H^2)^{n - 1}\right]\,,\\
\Omega_r &= X\left[1 + (1 - 2n)\alpha(6H^2)^{n - 1}\right]\,,\\
\Omega_{DE} &= -(1 - 2n)\alpha(6H^2)^{n - 1}\,.\label{eqn:denspardeplm}
\end{align}
\end{subequations}
We further calculate the deceleration parameter, which can be read off from the acceleration~\eqref{eqn:exaccel} and is given by
\begin{equation}
q = -1 + \frac{X + 3}{2}\frac{6H^2 + (1 - 2n)\alpha(6H^2)^n}{6H^2 + n(1 - 2n)\alpha(6H^2)^n}\,.
\end{equation}
We do not attempt to fit the parameters \(\alpha\) and \(n\) of the power law model based on the observational properties derived generically in this article, since this particular class of models has already been extensively studied and numerous numerical fits have been obtained~\cite{Bengochea:2008gz,Wu:2010mn,Bengochea:2010sg,Wei:2011mq,Nesseris:2013jea,Geng:2015hen,Basilakos:2016xob,Nunes:2016qyp,Oikonomou:2016jjh,Nunes:2016plz,Malekjani:2016mtm,Farrugia:2016qqe,Capozziello:2017bxm}. Instead we only give a qualitative estimate based on the phase diagrams shown in figure~\ref{fig:powerphasp}, again with the purpose to illustrate the generic formalism developed in this article. If one assumes that the qualitative behavior of the Hubble parameter is described by a Big Bang singularity \(H \to \infty\) at a finite time in the past, followed by an expansion, that finally leads to an accelerated expansion at a de Sitter fixed point, one is led to the conclusion that the expansion history of the universe is best described by either of the phase diagrams~\ref{fig:powerphasp1} or~\ref{fig:powerphasp5}. Hence, one concludes that the model parameters must satisfy either \(\alpha < 0\) and \(n < \frac{1}{2}\) or \(\alpha > 0\) and \(\frac{1}{2} < n < 1\). Remarkably, both of these possibilities are consistent with a positive density parameter \(\Omega_{DE}\), as can be seen from equation~\eqref{eqn:denspardeplm}.

This concludes our discussion of the power law model. We have seen that our general formalism reproduces a large number of results which have been previously obtained in individual studies. These findings demonstrate the validity and usefulness of our formalism.

\section{Conclusion}\label{ses:conclusion}
In this article we have derived a two-dimensional dynamical system from the flat FLRW cosmological field equations of a generic \(f(T)\) gravity theory, where the matter content is given by a combination of dust and radiation. We have shown that the full cosmological dynamics of this model depend only on a single function \(W(H)\) of the Hubble parameter \(H\), which is derived from the function \(f(T)\) defining the particular theory under consideration. Instead of choosing a particular form of \(f(T)\), we have kept the function fully generic and derived a number of physically relevant properties of the whole family of \(f(T)\) theories.

Our main result is comprised of numerous conditions on the Friedmann function \(W(H)\), which determine the existence and stability of fixed points in the cosmological dynamics, the possibility of a bounce or turnaround, the existence and severity of finite time singularities, the existence of accelerating and decelerating phases of the expansion of the universe and transitions between them as well as the possibility of crossing the phantom divide. As a fully generic result, we have shown that there exist no periodic orbits in the phase space, and no oscillating universe solutions, independent of the choice of the function \(f(T)\). Further, we have shown how points on the phase space and the shape of the Friedmann function \(W(H)\) at these points can be related to observational cosmological parameters. Note that our chosen matter content manifestly satisfies all energy conditions, and that all features we discussed are direct consequences of the modified gravitational dynamics.

To illustrate our results and the general formalism, we have applied it to a generic power law model \(f(T) = T + \alpha(-T)^n\). We have shown how the dynamics on the physical phase space depend on the constant parameters \(\alpha\) and \(n\) of the model and displayed the phase diagrams for all qualitatively different values of these parameters. We have further characterized all possible trajectories in these phase spaces and their acceleration and effective dark energy. In particular, we have shown that it is not possible to dynamically cross the phantom divide \(w_{DE} = -1\) in these models. We have finally shown that there are no trajectories that start from an initial accelerating period (which would be interpreted as inflation), become decelerating, and finally transition back to an accelerating de Sitter phase.

The formalism and generic results derived in this article can now be applied to any particular $f(T)$ gravity theory or class of such theories, in order to get a systematic overview of its cosmological behavior. It is left for future works to scrutinize other models in a similar manner, finally arriving at a catalog of $f(T)$ theories, classified by the dynamical properties of their cosmologies. Our results further hint towards the possibility to reverse the line of investigation and to construct heuristic $f(T)$ models based on a set of desired cosmological features. Once a class of models or a parameter range with viable dynamical behaviors has been confirmed, it can be subjected to further studies by other methods, e.g., the evolution of perturbations, local gravitational constraints, etc.

Finally, one may also consider a more general class of modified teleparallel theories augmented with a non-minimally coupled scalar field~\cite{Geng:2011aj,Otalora:2013tba,Jarv:2015odu,Abedi:2015cya}, a Gauss-Bonnet term~\cite{Kofinas:2014owa,Gonzalez:2015sha}, a boundary term~\cite{Bahamonde:2015zma}, combinations of those~\cite{Bahamonde:2015hza,Bahamonde:2016kba}, or higher derivatives of the torsion scalar~\cite{Otalora:2016dxe}. Additionally one may consider actions which are not a function of the torsion scalar~\eqref{eqn:torsionscal}, but of different contractions of the torsion tensor~\cite{Hayashi:1979qx,Maluf:2011kf,Bahamonde:2017wwk}. It should be straightforward to generalize our formalism to such theories, and thus to use our results to determine their cosmological dynamics.

\begin{acknowledgments}
The authors were supported by the Estonian Ministry for Education and Science Institutional Research Support Project IUT02-27 and Startup Research Grant PUT790, as well as by the European Regional Development Fund through the Center of Excellence TK133 ``The Dark Side of the Universe''. The authors thank Martin Kr\v{s}\v{s}\'ak and Christian Pfeifer for helpful comments and discussions.
\end{acknowledgments}

\appendix

\section{Classification of phase space points}\label{app:index}
In order to provide a better overview of all conditions on the Friedmann function \(W\) listed in the main part of the article, we provide a graphical ordering scheme of all values \(H\) that belong to the compactified phase space \(\bar{\mathcal{P}}\) in figure~\ref{fig:index}. Table entries refer to the corresponding general statements detailed in sections~\ref{sec:dynsys} to~\ref{sec:observ}. If several statements apply simultaneously to the same phase space point \((H, X)\), they are listed with an ampersand (\&) character. If several statements apply to the same value of \(H\), but different values of \(X\), they are separated with a pipe (|) character.

\begin{sidewaysfigure}[p]
\begin{tikzpicture}[align=center]
\tikzstyle{condition} = [rectangle, fill=black!10, draw=black!50]
\node[condition] at (3,0) {$W = 0$};
\node[condition] at (6,0) {$W \to \infty$};
\node[condition] at (14,0) {$W > 0$};
\node[condition] at (11,-1) {$W_H = 0$};
\node[condition] at (16,-1) {$W_H \neq 0$};
\node[condition] at (19,-1) {$W_H \to \pm \infty$};
\node[condition] at (0.5,-2.5) {$H > 0$};
\node[condition] at (0.5,-4.5) {$H < 0$};
\node[condition] at (0.5,-8.5) {$H = 0$};
\node[condition] at (0.5,-12.5) {$H \to \pm\infty$};
\node at (3,-2.5) {expan. dS FP;\\\ref{stmt:fixreg}. \ref{it:fpregisoat} | \ref{stmt:fixreg}. \ref{it:fpregisosa}};
\node at (3,-4.5) {contr. dS FP;\\\ref{stmt:fixreg}. \ref{it:fpregisore} | \ref{stmt:fixreg}. \ref{it:fpregisosa}};
\node at (6,-2.5) {expan. sing. FP;\\\ref{stmt:fixsing}. \ref{it:fpsingisore} | \ref{stmt:fixsing}. \ref{it:fpsingisosa}};
\node at (6,-4.5) {contr. sing. FP;\\\ref{stmt:fixsing}. \ref{it:fpsingisoat} | \ref{stmt:fixsing}. \ref{it:fpsingisosa}};
\node at (11,-3.5) {type~\ref{it:singtype2} sing.;\\\ref{stmt:singfi}. \ref{it:singfinz}};
\node at (16,-3.5) {$\dot{H} \neq 0$};
\node at (19,-3.5) {pseudo FP,\\type~\ref{it:singtype4} sing.;\\\ref{stmt:fixsing}. \ref{it:fpsingidnz} \& \ref{stmt:singff}. \ref{it:singffnz}};
\node at (3,-8.5) {static FP;\\\ref{stmt:fixreg}. \ref{it:fpregidwz}};
\node at (6,-8.5) {static sing. FP;\\\ref{stmt:fixsing}. \ref{it:fpsingidwd}};
\node[condition] at (9,-6.5) {$W_{HH} = 0$};
\node[condition] at (9,-8.5) {$W_{HH} \neq 0$};
\node[condition] at (9,-10.5) {$W_{HH} \to \pm\infty$};
\node at (12.5,-6.5) {type~\ref{it:singtype2} sing.;\\\ref{stmt:singfi}. \ref{it:singfiz}};
\node at (12.5,-8.5) {bounce / turnaround;\\\ref{stmt:bounce}};
\node at (12.5,-10.5) {static FP, type~\ref{it:singtype4} sing.;\\\ref{stmt:fixreg}. \ref{it:fpregidwhz} \& \ref{stmt:singff}. \ref{it:singffz}};
\node at (16,-8.5) {static FP;\\\ref{stmt:fixreg}. \ref{it:fpregnisoat} | \ref{stmt:fixreg}. \ref{it:fpregnisore}};
\node at (19,-8.5) {static FP;\\\ref{stmt:fixreg}. \ref{it:fpregidwhd}};
\node[condition] at (6.25,-12) {$(\ln W)_H \to 0$};
\node[condition] at (15.75,-12) {$(\ln W)_H \nrightarrow 0$};
\node at (6.25,-13) {type~\ref{it:singtype1} or type~\ref{it:singtype3} sing.; \ref{stmt:singii}};
\node at (15.75,-13) {inf. time sing.};
\draw (-0.5,-1.5) to (20.5,-1.5);
\draw (-0.5,-3.5) to (7.5,-3.5);
\draw (-0.5,-5.5) to (20.5,-5.5);
\draw (-0.5,-11.5) to (20.5,-11.5);
\draw (-0.5,-13.5) to (20.5,-13.5);
\draw (1.5,0.5) to (1.5,-13.5);
\draw (4.5,0.5) to (4.5,-11.5);
\draw (7.5,0.5) to (7.5,-11.5);
\draw (20.5,0.5) to (20.5,-13.5);
\draw[dashed] (7.5,-7.5) to (14.5,-7.5);
\draw[dashed] (7.5,-9.5) to (14.5,-9.5);
\draw[dashed] (14.5,-0.5) to (14.5,-11.5);
\draw[dashed] (17.5,-0.5) to (17.5,-11.5);
\draw[dashed] (11,-11.5) to (11,-13.5);
\end{tikzpicture}
\caption{Classification of all points in the compactified physical phase space \(\bar{\mathcal{P}}\). Gray fields indicate conditions, while white fields show the physical consequence if all conditions in the same row and column are satisfied. Numbers and codes refer to the statements and cases in the main part of the article. FP = fixed point, dS = de Sitter.}
\label{fig:index}
\end{sidewaysfigure}

\bibliography{ftcosdyn}
\end{document}